\documentclass[12pt]{article}
\usepackage[left=2cm, right=2cm, bottom=2cm ,top=2cm, footskip=1cm]{geometry}
\usepackage[font=small,labelfont=bf]{caption}

\usepackage{sectsty} 
\subsubsectionfont{\normalfont\itshape}

\usepackage{setspace}
\usepackage{chngcntr}

\usepackage{hyperref}

\usepackage[authoryear,sort&compress]{natbib}
\setcitestyle{open={(},close={)}}

\usepackage{graphicx}
\usepackage{amsmath}

\usepackage[section]{placeins}

\usepackage[table,dvipsnames]{xcolor}
\usepackage{multirow}
\usepackage{makecell}

\usepackage{subcaption} 
\usepackage{float} 

\definecolor{delta}{HTML}{8CA9FF}
\definecolor{theta}{HTML}{CC9DFC}
\definecolor{alpha}{HTML}{FC7EED}
\definecolor{beta}{HTML}{ADFC90}
\definecolor{gamma}{HTML}{82E8E8}

\title{Multi-centre normative brain mapping of intracranial EEG lifespan patterns in the human brain}
\author{Heather Woodhouse$^{1}$, Gerard Hall$^{1}$, Callum Simpson$^{1}$, Csaba Kozma$^{1}$,\\Frances Turner$^{1}$, Gabrielle M. Schroeder$^{1,O}$, Beate Diehl$^{3}$, John S. Duncan$^{3}$,\\Jiajie Mo$^{4}$, Kai Zhang$^{4}$, Aswin Chari$^{5}$, Martin Tisdall$^{5}$, Friederike Moeller${^5}$,\\Chris Petkov$^{2,6}$, Matthew A. Howard$^{6}$, George M. Ibrahim$^{7}$, Elizabeth Donner$^{7}$,\\Nebras M. Warsi$^{7}$, Raheel Ahmed$^{8}$, Peter N. Taylor$^{1,2,3}$,  Yujiang Wang$^{1,2,3*}$}
\date{}

\begin{document}
\pagenumbering{gobble} 

\newgeometry{left=2cm, right=2cm, bottom=2cm ,top=0.5cm, footskip=1cm}

\maketitle
\thispagestyle{empty}
\begin{small}
\begin{enumerate}
\item{CNNP Lab (www.cnnp-lab.com), Interdisciplinary Computing and Complex BioSystems Group, School of Computing, Newcastle University, Newcastle upon Tyne, United Kingdom}
\item{Faculty of Medical Sciences, Newcastle University, Newcastle upon Tyne, United Kingdom}
\item{UCL Queen Square Institute of Neurology, Queen Square, London, United Kingdom}
\item{Beijing Tiantan Hospital, Beijing, China}
\item{Great Ormond Street Hospital for Children, London, United Kingdom}
\item{University of Iowa Hospitals and Clinics, Iowa City, IA, United States}
\item{The Hospital for Sick Children, University of Toronto, Toronto, Canada}
\item{University of Wisconsin-Madison, Madison, WI, United States}
\end{enumerate}

\begin{center}
* Yujiang.Wang@newcastle.ac.uk \hspace{1cm} $^O$ G.M.S. ORCID ID: 0000-0003-2278-5227
\end{center}
\end{small}

\restoregeometry



\begin{abstract}

\textbf{Background:} Understanding healthy human brain function is crucial to identify and map pathological tissue within it. Whilst previous studies have mapped intracranial EEG (icEEG) from non-epileptogenic brain regions, these maps do not consider the effects of age and sex. Further, most existing work on icEEG has often suffered from a small sample size due to the modality’s invasive nature. Here, we substantially increase the subject sample size compared to existing literature, to create a multi-centre, normative map of brain activity which additionally considers the effects of age, sex and recording hospital.

\textbf{Methods:} Using interictal icEEG recordings from $n = 502$ subjects originating from 15 centres, we constructed a normative map of non-pathological brain activity by regressing age and sex on relative band power in five frequency bands, whilst accounting for the hospital effect.

\textbf{Results:} Recording hospital significantly impacted normative icEEG maps in all frequency bands, and age was a more influential predictor of band power than sex. The age effect varied by frequency band, but no spatial patterns were observed at the region-specific level. Certainty about regression coefficients was also frequency band specific and moderately impacted by sample size. 

\textbf{Conclusion:} The concept of a normative map is well-established in neuroscience research and particularly relevant to the icEEG modality, which does not allow healthy control baselines. Our key results regarding the hospital site and age effect guide future work utilising normative maps in icEEG.

\end{abstract}

\newpage
\pagenumbering{arabic} 

\section{Introduction}

Age and sex are important factors which are known to influence brain activity. Understanding \textit{how} these variables affect the brain is important for both clinical applications and academic research. Using scalp EEG recordings in children and adolescents, past studies found that age has a negative relationship with relative power in slower frequency bands, namely $\delta$ and $\theta$, but a positive relationship with faster ones ($\alpha$, $\beta$) \citep{Gasser1988, Clarke2001}. Both MEG and scalp EEG studies report some sex differences in the same age range, with males tending to have more $\alpha$-power \citep{Ott2021, Clarke2001}. A study in MEG across the whole lifespan found similar frequency band-specific relationships between age and power, whilst a young adult study in scalp EEG reported various sex differences during the resting state \citep{Cave2021, Gómez2013}. In summary, whilst consistent effects across the lifespan have been reported, a conclusive map does not currently exist for electrical brain activity as it does for structural neuroimaging \citep{Bethlehem2022}.

Although much work has been done to assess the impact of age and sex on MRI, MEG and scalp EEG, to our knowledge intracranial EEG (icEEG) has not yet been investigated in this context, most likely because its invasive nature precludes data collection from healthy controls \citep{Gomez2017, Hinault2022, CamCan2014, Coffey1998}. To overcome this issue, researchers have collected icEEG recordings from individuals with epilepsy, from brain regions that were later deemed not pathological and not epileptogenic. By combining these recordings over many individuals, a map of normative brain activity has been proposed \citep{Groppe2013, Frauscher2018, Taylor22}. Whilst there has been exciting research into normative maps using icEEG and their potential for epilepsy research, the effect of age and sex on icEEG has not been determined \citep{Taylor22, Frauscher2018, Kalamangalam2020, Wang2023, Bernabei2022}.

To analyse this effect thoroughly, this investigation utilises a large data set, encompassing multiple international hospitals. This is similar to \cite{Bethlehem2022} which leverages MRI scans from various studies globally to create brain charts for the human lifespan -- our study seeks to help establish such methods in the icEEG literature. Importantly, we will perform our analysis on the largest multi-centre normative icEEG dataset to date ($n=502$ subjects), accounting for hospital site effects in data by using mixed-effect modelling. Following this, we aim to uncover and discuss the relationships, if any, between band power extracted from icEEG recordings, and the variables age, sex and recording hospital. 

Ultimately, we highlight the need to account for the heterogeneity of icEEG data. We seek to fill a gap in the literature regarding the effect of two standard covariates -- age and sex -- on band power values extracted from icEEG and modelled in a normative setting. We also expand on previous sample sizes and properly consider hospital variability. Such research is necessary, as an understanding of ageing patterns, sex differences and hospital variation in this modality will conceivably aid in the identification of pathological activity through deviations from `normal' trends. 

\section{Methods}

\subsection{Subjects}

Our analysis involved 502 individuals with epilepsy undergoing presurgical evaluation with icEEG to localise the seizure onset zone. Data was collected from Beijing Tiantan Hospital, Great Ormond Street Hospital, University of Iowa Hospital, SickKids, University College London Hospital and the University of Wisconsin-Madison. We also used nine hospitals from the publicly available RAM data set (https://memory.psych.upenn.edu/RAM), bringing our total number of contributing hospital sites to 15. 

\begin{figure}[ht]
    \centering
\includegraphics[width=\textwidth]{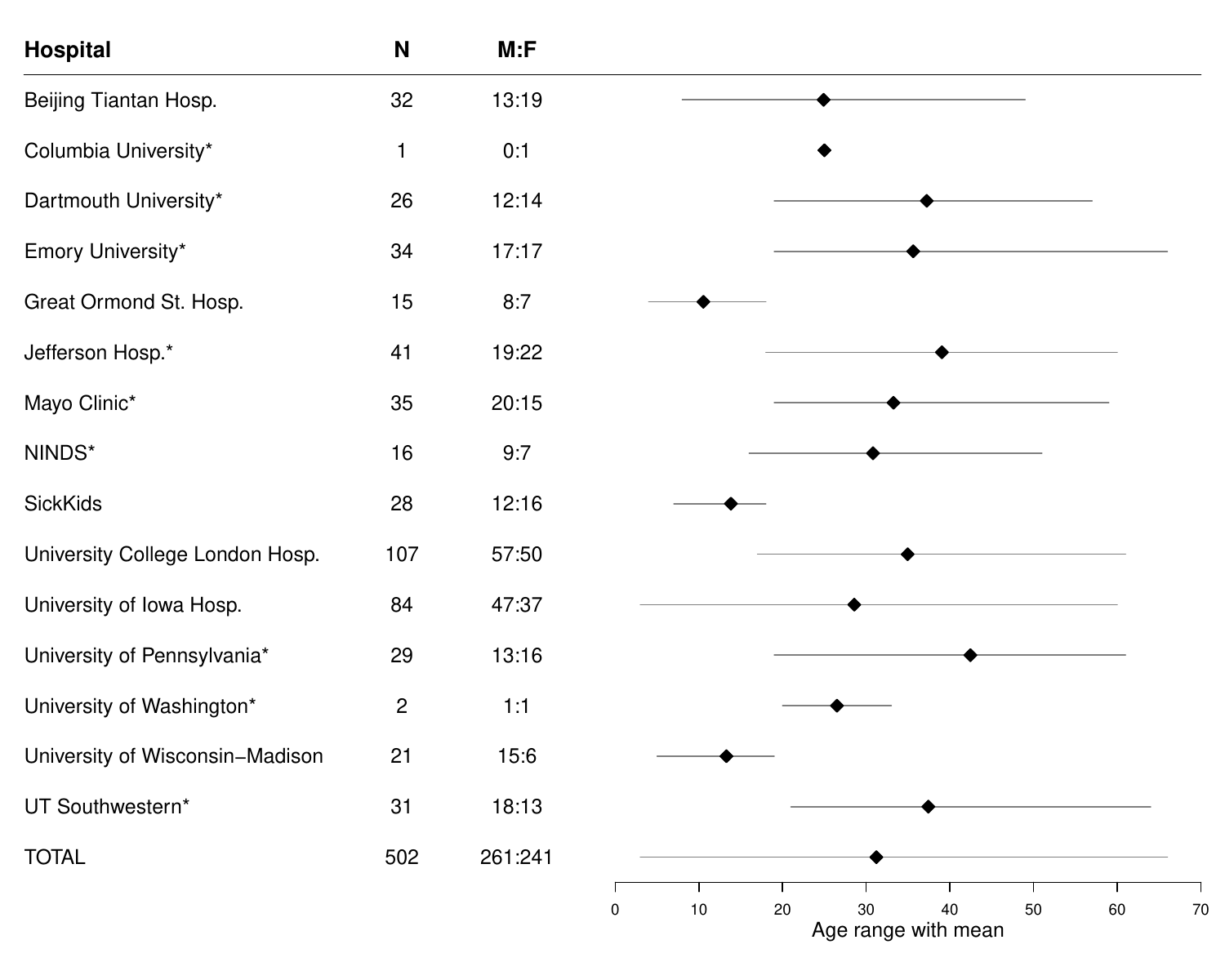}
    \caption{The number of subjects from each participating hospital, along with age and sex distributions. Hospitals from the RAM cohort are highlighted with an *.}
    \label{F:EDA}
\end{figure}

Anonymised icEEG recordings were analysed following approval of the Newcastle University Ethics Committee (reference number 23973/2022). Both grid and depth electrode recordings were included. A summary of our final cohort is provided in Figure~\ref{F:EDA}.

\subsection{Electrode localisation}

For every participant, electrode contacts were localised to regions of interest (ROIs) according to a predefined parcellation, using the ``Lausanne scale 36" atlas, with 82 ROIs \citep{Hagmann2008}. The Lausanne atlas has been used previously for normative intracranial analysis \citep{Betzel2019, Taylor22}.

The methods for localising the electrode contacts to brain regions have also been described previously \citep{Taylor22, Wang2020, Hamilton2017, Wang2023}. Different hospital sites provided different levels of data, so our methods for localising electrode contacts to ROIs varied between them. Hospitals either provided contact locations in MNI space, or provided native space imaging and co-localised contacts. In the first case, we assigned electrodes to one of 82 regions from the Lausanne scale 36 atlas \citep{Hagmann2008}. We used FreeSurfer to generate volumetric parcellations of an MNI space template brain \citep{FreeSurfer2012, Hagmann2008}. Each electrode contact was assigned to the closest grey matter volumetric region within 5~mm. If the closest grey matter region was $>$5~mm away then the contact was excluded from further analysis. For the latter case, a similar technique was used, but applied in native space using the subject's own parcellated pre-operative MRI. 

To ensure our findings were robust to parcellation choice, we confirmed that key results held using a finer-grained parcellation (Supplementary \ref{S:ParcellationValid}).

\subsection{icEEG processing}

\subsubsection*{Segment selection}

For each subject, we extracted a 70-second interictal icEEG segment from a period of relaxed wakefulness, at least two hours away from any detected ictal events. The raw signals and the power spectral densities were visually and algorithmically inspected for spikes, artefacts and faulty channels (see Supplementary \ref{S:BadChannels} for details). Using clinical reports (if available) we excluded any contacts that were: within known lesions, within the seizure onset zone, or subsequently resected. This ensured that at the subject level, we only retained channels thought to be non-pathological in terms of both location in the brain and the signal produced.

If unavailable in clinical reports, information on resected contacts could also be obtained by drawing masks using  post- and pre-operative scans (where available), as in \cite{Taylor22} and \cite{Wang2020}. Supplementary \ref{S:ChannelDetails} details the exclusionary information available for each hospital. 

\subsubsection*{Signal processing}

All segments were bandpass filtered between 0.5-80 Hz and downsampled to 200 Hz with an anti-alias filter. A common average reference was applied to all recordings and the power spectral density was calculated using Welch's method with a 2-second window and 1-second overlap. The average band power was then calculated in five frequency bands: $\delta$ (1-4 Hz), $\theta$ (4-8 Hz), $\alpha$ (8-13 Hz), $\beta$ (13-30 Hz) and $\gamma$ (30-77.5 Hz). 

In the $\gamma$-band, data between 47.5-52.5 Hz and 57.5-62.5 Hz were excluded from all hospital sites, to avoid any power line noise. The $\gamma$-band was also capped at 77.5 Hz due to 80 Hz noise in the RAM cohort. Band power estimates were log$_{10}$ transformed and normalised to sum to one in each contact (L1-normalised). These final values are used throughout results to represent relative band power, denoted RBP($\cdot$).

\subsection{Normative data table creation}
\label{M:pooling}

At this stage, the electrode contacts from each subject have been assigned to a single, nearest ROI and the RBP($\cdot$) in five frequency bands has been computed. If multiple contacts from one subject were assigned to the same ROI, then the RBP($\cdot$) values in each frequency band were averaged across contacts to obtain single values of RBP($\cdot$) per region per subject. 

We excluded six sub-cortical regions due to no, or a very low number of samples: pallidum, thalamus and accumbens area in both hemispheres, reducing our total number of ROIs from 82 to 76. All cortical regions were retained.

Previous work has demonstrated a left/right symmetry in RBP($\cdot$) in EEG, MEG, and icEEG \citep{Taylor22, Janiukstyte23, Owen2023}. Hence, to maximise the number of subjects in each region, we created a `mirrored' version of our data table, in which, the data from homologous regions were combined (e.g., the left amygdala and the right amygdala).  In the case of bilateral implantation of symmetric regions, we ensured individuals only had one value of RBP($\cdot$) per ROI and frequency band. This mirrored data table only comprised 38 regions but had markedly more subjects in each. Throughout the results, it will be made clear whether the mirrored (regional), or original (whole-brain), normative data table is in use at any given time.

We confirm the validity of this mirroring in Supplementary \ref{S:PoolingValid} by repeating one of our key results using the original data. Additionally, Supplementary \ref{S:AgeOnset} demonstrates that our normative values are not related to a subject's age of epilepsy onset, a feature representing pathology. Unfortunately, we do not have the data to check other disorder-specific features, such as drug levels or epilepsy classification.

In summary, our final normative data table included a unique subject identifier, their age and sex, their originating hospital, and, for the regions in which they had electrode implantation only, their RBP($\cdot$) in five frequency bands averaged across contacts where necessary. 

\subsection{Statistical modelling and testing}
\label{M:optLMM}

The final step was to fit a suitable model to the normative data to examine the effect of age, sex and hospital site on relative band power. Visual inspection of scatter plots of the RBP($\cdot$) values against age in each frequency band (at the whole-brain level) showed no evidence of a non-linear relationship (see Supplementary \ref{S:LinearValid}). Further, the effect of having multiple recording hospitals had to be considered, especially as some hospitals supplied only paediatric, or only adult recordings. Therefore, we implemented a linear mixed model (LMM) in each frequency band, specifically a random intercept model, with the originating hospital site as a random effect. Our cohort has a wide range of subjects per hospital (Figure \ref{F:EDA}), but one of the strengths of mixed modelling is the ability to handle unequal group sizes \citep{West2014}.

Possible fixed effects were age and sex, meaning four LMMs were under consideration in each frequency band:
\begin{alignat*}{2}
  &  \textbf{Null:} \; && RBP(\cdot) \sim (1|Hospital) \\
  &  \textbf{Age:} \; && RBP(\cdot) \sim Age +(1|Hospital) \\
  &  \textbf{Sex:} \; && RBP(\cdot) \sim Sex +(1|Hospital) \\
  &  \textbf{Full:} \; && RBP(\cdot) \sim Age + Sex +(1|Hospital)
\end{alignat*}

An interaction term between age and sex was also considered but deemed unnecessary (Supplementary \ref{S:interaction}). Model fitting could be performed at either the regional or whole-brain level. Taking the latter, we calculated several model evaluation statistics to determine the optimal fixed effect structure of the LMM in each frequency band. Since we worked at the whole-brain level, the original data table was used. The statistics under consideration were the AIC and BIC for each model, 95\% profiled confidence intervals for the regression coefficients of fixed effects in all but the null model (which has none) and the likelihood ratio test $p$-values for every pair of models which differed by one variable. 

In four frequency bands, each metric was in agreement regarding the optimal LMM, so no further tests were required. The preferred models were the age model in the $\delta$ and $\beta$ bands; the sex model in $\gamma$ and the full model in $\theta$. The $\alpha$ band demonstrated some uncertainty between the age model and the full model. Since both models perform similarly, we retain the simpler age model as the optimal choice in the $\alpha$ band. 

\begin{figure}[ht]
    \centering
    \includegraphics[width=\textwidth]{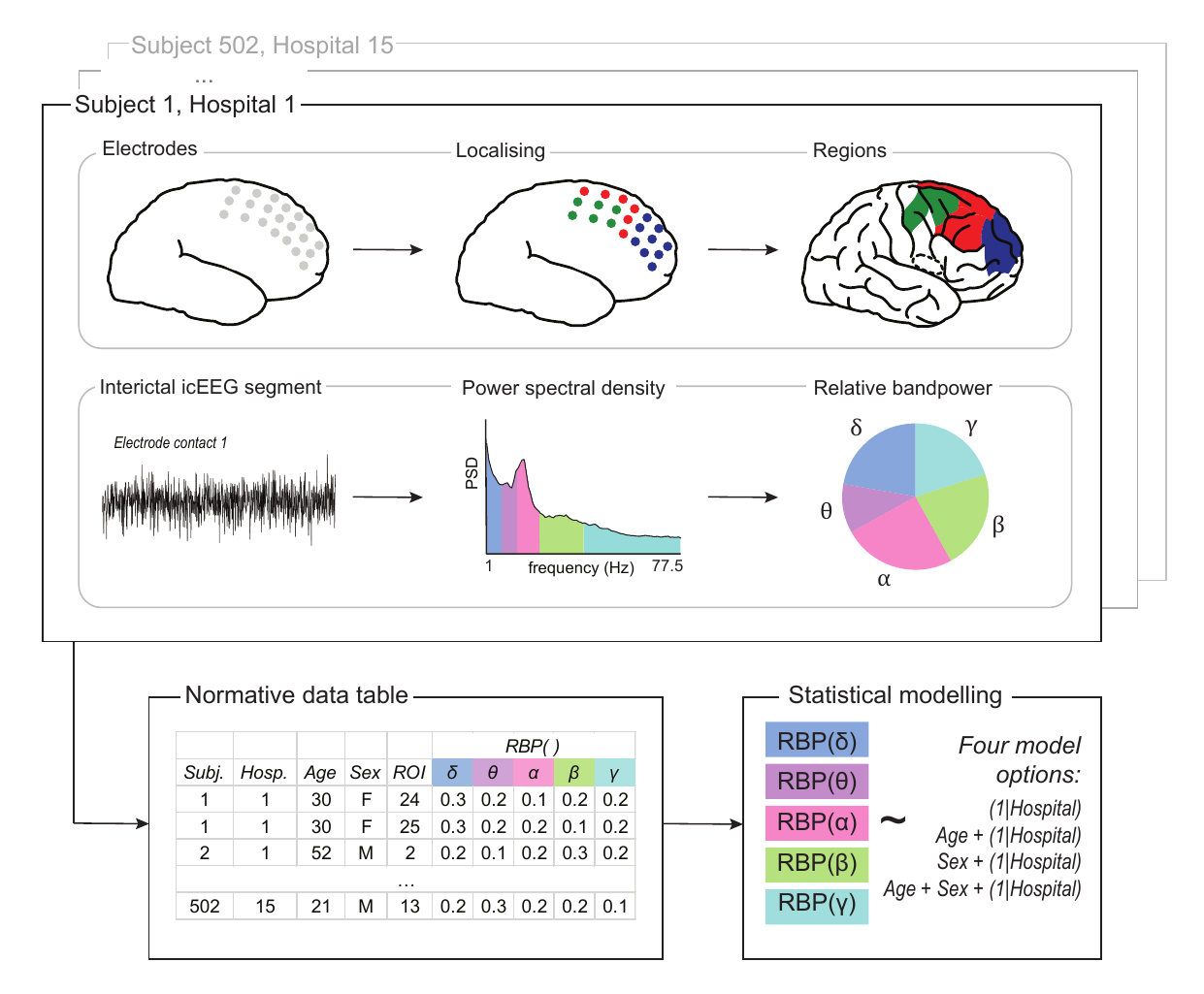}
    \caption{A visual representation of our methods: electrode localisation (top) and signal processing following segment selection (middle), for each subject. Subjects were combined to create our normative data table, to which we applied our statistical models (bottom). All components show dummy data for example purposes only. Brain plots from \cite{BrainPlots}.}
    \label{F:methods}
\end{figure}

Hence, the optimal LMM was frequency band-specific in our cohort, and so we report frequency band-specific effects throughout our results. We have provided a brief visual overview of our methods (Figure~\ref{F:methods}). In the results section, we explore both random (hospital site) and fixed (age, sex) effects. Specifically, we investigate if the variables explain RBP($\cdot$) variation, and examine how they might influence it.

\section{Results}

Results \ref{R:hospitalEffect}, \ref{R:AgeSexWeak} and \ref{R:AgeModelBest} examine LMMs at the whole-brain level, using the original data. Subsequent results additionally consider the regional-level analysis, and therefore incorporate the mirrored data, to determine if any spatial variations are present in our findings. 

\subsection{Hospital site effects impacted icEEG normative maps}
\label{R:hospitalEffect}

Focusing first on the random effect structure (recording hospital), we found that in our cohort and model, the hospital site effect was much more powerful in explaining RBP($\cdot$) differences at the whole-brain level than any fixed effect structure, in almost all cases. Table \ref{T:ICCR2} quantifies this, showing marginal $R^2$ values ($R^2_m$) and intraclass correlation coefficients ($ICC$) as percentages for the full model, the age model and the sex model. The $R^2_m$ represents the proportion of variation in RBP($\cdot$) explained by the fixed effect(s) of that model. The $ICC$ represents the proportion of the variance explained by the recording hospitals. The optimal LMM as per Section \ref{M:optLMM} has been highlighted for each frequency band.

Note, the $R_m^2$ values of the two single fixed effect models do not sum exactly to the $R^2_m$ of the full model, because the denominator in the calculation of $R_m^2$ includes the residual variance and therefore changes with each model.

\begin{table}[hb]
    \setlength{\tabcolsep}{12pt}
    \centering
    \begin{tabular}{|c|cc|cc|cc|}
    \cline{2-7}
    \multicolumn{1}{c|}{} & \multicolumn{2}{c|}{Full model} & \multicolumn{2}{c|}{Age model} & \multicolumn{2}{c|}{Sex model}\\
    \multicolumn{1}{c|}{} & $R^2_m$  & $ICC$ & $R^2_m$ & $ICC$ & $R^2_m$ & $ICC$ \\
    \hline
       $\delta$  &  4.28  & 15.23  & \cellcolor{delta}4.27 & \cellcolor{delta}15.27  & 0.00 & 19.42 \\
       $\theta$  &  \cellcolor{theta}0.60  & \cellcolor{theta}15.17 & 0.23 & 14.88 & 0.41 & 16.25 \\
       $\alpha$  &  8.04  & 5.09  & \cellcolor{alpha}7.97 & \cellcolor{alpha}5.13  & 0.00 & 5.88  \\
       $\beta$   & 2.59  & 6.58  & \cellcolor{beta}2.58 & \cellcolor{beta}6.58  & 0.00 & 7.93  \\
       $\gamma$  &  0.25  & 32.59 & 0.01 & 32.16 &  \cellcolor{gamma}0.24 & \cellcolor{gamma}32.25 \\
    \hline
    \end{tabular}
    \caption{$R^2_m$ and $ICC$ values (measured in \%) for the full model, the age model and the sex model. $R^2_m$ represents the proportion of variation in RBP($\cdot$) explained by the fixed effect(s) of that model. The $ICC$ represents the proportion of the variance explained by the grouping structure, namely recording hospitals. The optimal covariate subset, as determined by a standard model selection process, is highlighted for each frequency band.}
    \label{T:ICCR2}
\end{table}

There is striking variation in the magnitude of the hospital site effect on RBP($\cdot$) across signal properties. However the $ICC$s consistently exceed 5\%, indicating that, in our cohort, at least a twentieth of the variation in RBP($\cdot$) was explained by the random effect structure. This proportion sometimes reached as high as 30\%, attributing that recording hospital impacted RBP($\cdot$) in all frequency bands. Further, in all bands except $\alpha$, the $ICC$ was consistently larger than the $R^2_m$ values across models, with this difference being most notable in the $\theta$ and $\gamma$ bands.

\begin{figure}[ht]
    \centering
    \includegraphics[width=\textwidth]{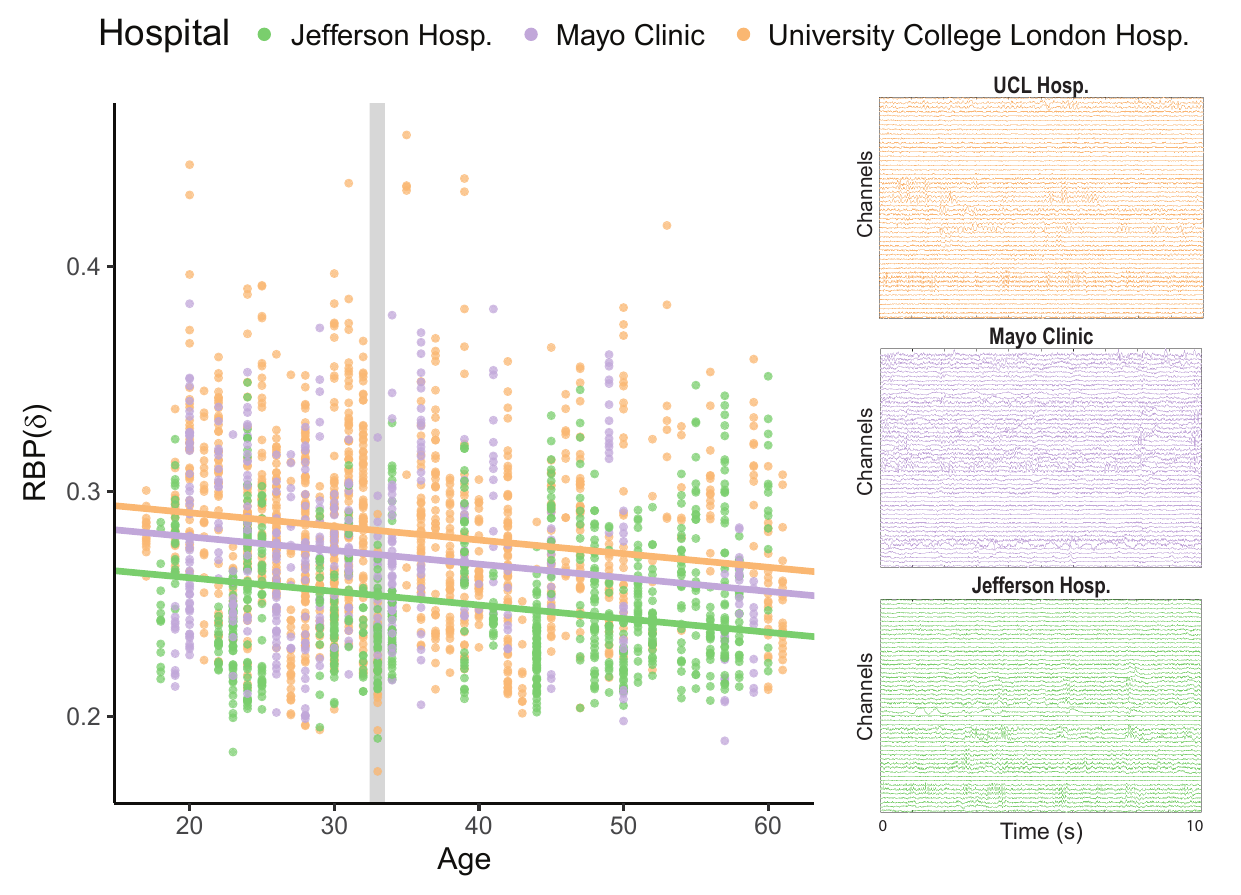}
    \caption{Scatterplot of RBP($\delta$) against age for thee well-populated hospitals. The age model was implemented for all hospitals, then the age slope and intercepts for Jefferson Hospital, Mayo Clinic and University College London Hospital were extracted and plotted. For three similar individuals, men of age 33, the first 50 channels and 10 seconds of their 70-second preprocessed icEEG segments have been provided. The grey bar indicates where the segments from the 33-year-old men lie within this subset of the data.}
    \label{F:hosp_effects}
\end{figure}

To visualise the effect of the originating hospital on RBP($\cdot$), we fit the age model to all data, then selected three well-populated hospitals with similar age ranges and plotted RBP($\delta$) against age, along with the model fit in Figure \ref{F:hosp_effects}. Additionally, we identified a 33-year-old male from each of the three hospitals and have provided the first 10 seconds of their processed 70-second icEEG segment.  The three subjects had variable numbers of channels, so only the first 50 channels are displayed for comparative purposes. Figure \ref{F:hosp_effects} demonstrates that while individuals' icEEG segments may appear similar, the originating hospital introduces underlying differences to relative band power properties. Visualisations for all five frequency bands and all hospitals can be found in Supplementary \ref{S:HospEffects}. 

\subsection{Age and sex explained only some of the variation in relative band power}
\label{R:AgeSexWeak}

We next examined our fixed effects at the whole-brain level ($R^2_m$). Neither age nor sex explained a substantial portion of the variance in the RBP($\cdot$), in any frequency band (Table \ref{T:ICCR2}). The largest effect was found in the $\alpha$-band, where $R^2_m$ was highest at 8\%, which in contrast, is similar to the minimum $ICC$ values at around 5\%.

Across fixed effects and frequency bands, all $R^2_m$ values were below 10\%, with over half being less than 1\%. This suggests that other factors significantly affect RBP($\cdot$) on the whole-brain level, which are not considered here.

\subsection{Age was more important than sex for predicting relative band power}
\label{R:AgeModelBest}

Comparing $R^2_m$ for the age and sex models revealed which of the two was more valuable when predicting RBP($\cdot$). All bands which achieve $R^2_m>1\%$ do so only when age is included in the model (Table \ref{T:ICCR2}). In fact, sex consistently accounted for a negligible portion of the variance in the response, with $R^2_m$ values below 0.5\%. This statement holds even in $\theta$ and $\gamma$, with sex retained in the optimal LMM, confirming sex was not a significant predictor in this cohort. In contrast, $R^2_m$ values for age fluctuated notably across frequency bands, being higher in $\delta$, $\alpha$ and $\beta$ ($R^2_m$ range of $7.96\%$ and $0.41\%$ for age and sex models respectively). 

\begin{figure}[!h]
    \centering
    \includegraphics[width=\linewidth]{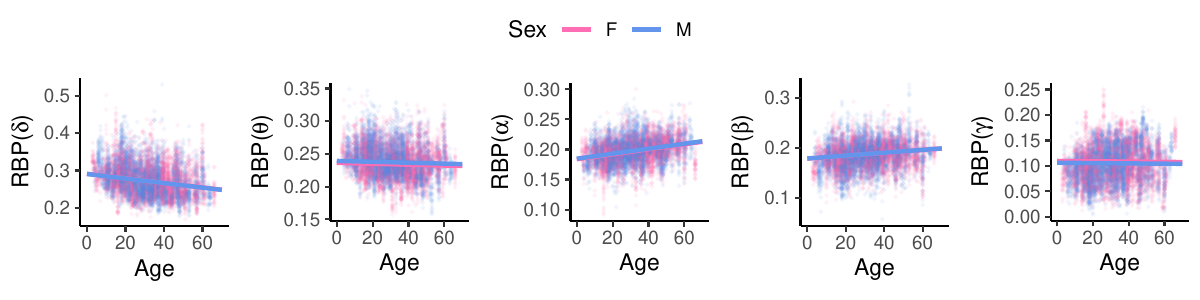}
    \caption{Scatterplots of RBP($\cdot$) against age in every frequency band, using data at the whole-brain level. The full model has been implemented and the male/female regression lines have been plotted.}
    \label{F:sex_effect}
\end{figure}

To validate the lack of sex effect in this cohort, we plotted RBP($\cdot$) against age for all frequency bands at the whole brain level. We then fit the full model and plotted the male/female regression lines. Figure \ref{F:sex_effect} demonstrates visually that sex is not a significant predictor in this cohort, with the differing sex lines being indistinguishable in almost all bands. 

Together, Table \ref{T:ICCR2} and Figure \ref{F:sex_effect} emphasize that the influence of sex was minor, and age was the more important predictor of RBP($\cdot$) in our cohort. Hence, subsequent analysis focuses on the age model for all frequency bands; however, we exercise caution when interpreting results in the $\theta$ and $\gamma$ bands, which have a weak relationship with both fixed effects.

\subsection{The effect of age on band power was spatially uniform and frequency band dependent}
\label{R:age&space}

With the focus shifted to the age model alone, our goal was to understand \textit{how} age impacted RBP($\cdot$). 

\begin{figure}[!h]
    \centering
    \includegraphics[width=\textwidth]{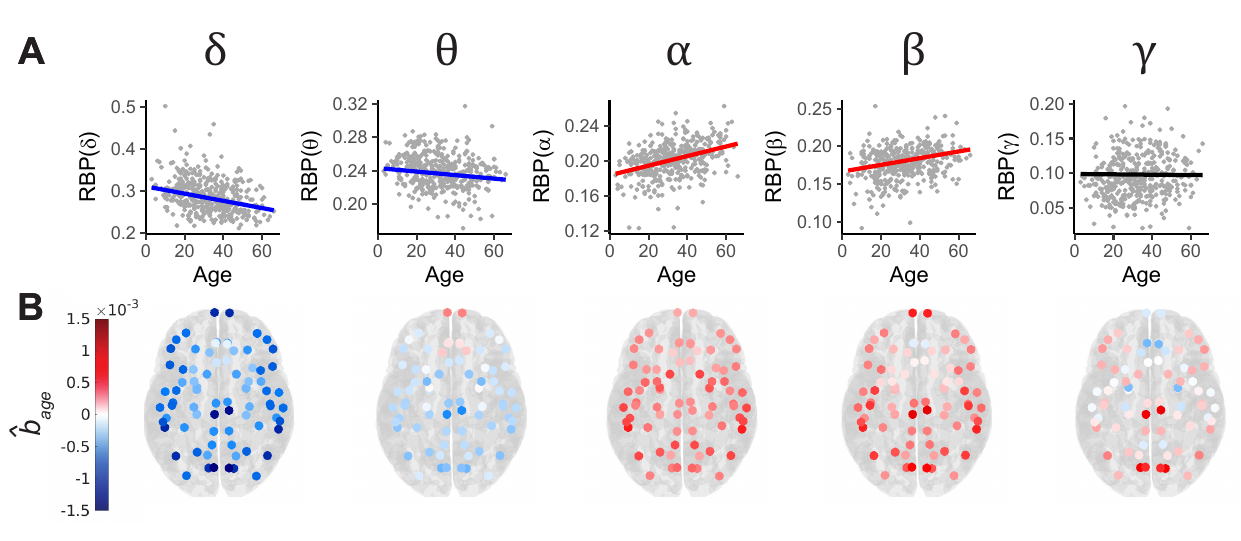}
    \caption{\textbf{A)} Visualisations of the age model in each frequency band in the most densely populated region, the middle temporal. Using the 95\% confidence intervals for $\hat{b}_{age}$ in the region, blue regression lines indicate a negative relationship between RBP($\cdot$) and age, whilst red regression lines indicate a positive one and black indicates no relationship. \textbf{B)} Values of $\hat{b}_{age}$ from the age model at the region-level. Values are shown for each ROI and each frequency band of interest. The colour scale is symmetric and fixed across frequency bands with blue representing negative regression coefficients and red representing positive ones. Data from symmetric regions were mirrored, i.e. results are reflected across the midline to provide a whole-brain visualisation.}
    \label{F:brains}
\end{figure}

Taking the most densely populated ROI (the middle temporal region), the age model was implemented for RBP($\cdot$) in each frequency band and regression coefficients for age (denoted $\hat{b}_{age}$) were extracted. A visualisation of model fit in this region is provided in Figure \ref{F:brains}A, where blue coloured lines indicate that RBP($\cdot$) decreased with age, red lines indicate an increase with age and black lines indicate no relationship (based on 95\% confidence intervals for $\hat{b}_{age}$). This method can be extended to all regions, extracting $\hat{b}_{age}$ for each frequency band and region. The $\hat{b}_{age}$ values can then be visualised on the brain using a hot/cold colour scale, indicating increasing/decreasing RBP($\cdot$) with age (Figure \ref{F:brains}B). Finally, disregarding regional information, at the whole-brain level, the age model was implemented in each frequency band and regression coefficients were extracted, along with their 95\% confidence intervals (Table \ref{T:b_age_CIs}). Looking at both whole-brain and ROI-level results allowed us to determine if the results had any spatial variation.

\begin{table}[!h]
    \centering
    \begin{tabular}{|c|c|c|c|c|}
        \hline
         $\delta$ & $\theta$ & $\alpha$ & $\beta$ & $\gamma$ \\
        \hline
        $-6.1 \times 10^4$ & $-0.8 \times 10^4$ & $4.1 \times 10^4$ & $2.9 \times 10^4$ & $-0.2 \times 10^4$ \\
         $(-6.7,-5.4) \times 10^4$ & $(-1.2,-0.4) \times 10^4$ & $(3.8,4.5) \times 10^4$ & $(2.4,3.3) \times 10^4$ & $(-0.7,0.3) \times 10^4$ \\
        \hline
    \end{tabular}
    \caption{Values of $\hat{b}_{age}$ along with 95\% confidence intervals for each frequency band. Values are rounded to 5 decimal places and were calculated by applying the age model at the whole-brain level.}
    \label{T:b_age_CIs}
\end{table}

Opposing relationships between RBP($\cdot$) and age were observed at both the region-specific and whole-brain levels, with decreasing trends in $\delta$ and $\theta$ frequencies, increasing trends in the $\alpha$ and $\beta$ bands and weaker results found for the $\gamma$ band.

Figure~\ref{F:brains}B reveals no discernible spatial gradient in any frequency band, but a clear switch in the sign of $\hat{b}_{age}$ occurs between $\theta$ and $\alpha$, from negative to positive. The $\theta$ and $\gamma$ trends were notably weaker, presenting values of $\hat{b}_{age}$ which are closer to zero, along with some regions deviating from the overarching trends. This is consistent with Table \ref{T:ICCR2} showing that for these two bands, the relationship between age and RBP($\cdot$) is weak, when compared with $\delta$, $\alpha$ and $\beta$.

On the whole-brain level, 95\% confidence intervals on $\hat{b}_{age}$ reinforced previous results. Table \ref{T:b_age_CIs} provides no evidence for an age-RBP($\gamma$) relationship and $\theta$'s confidence interval is relatively near zero compared with the remaining three bands. These confidence intervals support the existence of a negative relationship between RBP($\delta$) and age, along with a positive one between RBP($\alpha$) and RBP($\beta$), and age. Hence, the effect of age on RBP($\cdot$) was undoubtedly frequency band-specific in this cohort.

The subset of subjects varies between ROIs in the regional analysis; however, the age distribution in each region did not drive any differences between them (Supplementary \ref{S:ROIages}).
The trends in Figure~\ref{F:brains}B persist in a finer-grained parcellation, and when using the original data in which hemispheres were not mirrored (Supplementary \ref{S:ParcellationValid} and \ref{S:PoolingValid}).

\subsection{Certainty about age effect was frequency band specific and impacted by sample size}
\label{R:samplesize}

\begin{figure}[hp]
    \centering
    \includegraphics[width=0.92\textwidth]{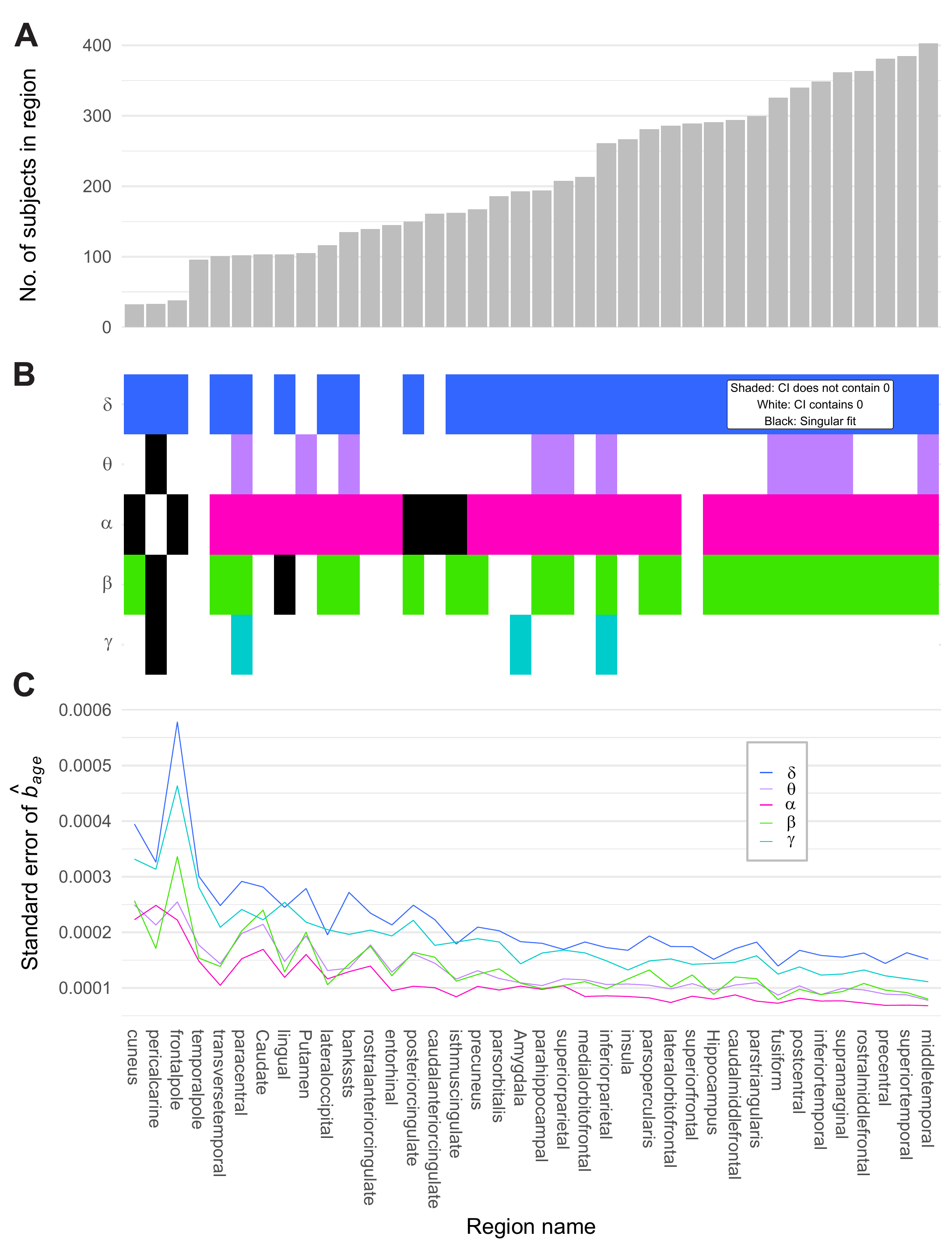}
    \caption{Summaries for each ROI under consideration. The $x$-axis displays the region name and is ordered by the number of subjects per region. \textbf{A}) number of subjects. \textbf{B}) a binary indicator of whether the 95\% confidence interval for $\hat{b}_{age}$ contains 0 in each frequency band. Singular fits given in black. \textbf{C}) standard errors of $\hat{b}_{age}$ in each frequency band.}
    \label{F:ROI_stats}
\end{figure}

Despite substantial data collection efforts, when analysing region-level models there was an impact of low sample size. Figure \ref{F:ROI_stats} displays region-level summaries in every frequency band: the standard error for $\hat{b}_{age}$, the number of subjects in the region, and a binary indicator of whether or not the 95\% confidence interval for $\hat{b}_{age}$ contains 0. It also highlights any regions where the model produced a singular fit. 

As a side note, extreme values in Figure \ref{F:brains}B co-localise to low sampled regions seen in Figure \ref{F:ROI_stats}, e.g., the frontal pole stands out from the general trend in $\theta$ and only has $n=38$, highlighting the importance of sample size.

The standard errors revealed that smaller sample sizes can lead to regression coefficient standard errors more than double those of the highly implanted ROIs. However, there appeared to be a threshold around $n=150$, beyond which the $\hat{b}_{age}$ standard error lines plateaued, suggesting a lower limit of $n$, beyond which we can be confident in our findings. 

The $\delta$ band displayed the highest regression coefficient standard errors almost consistently across ROIs, followed by $\gamma$ with slightly lower values. Comparatively, $\alpha$, $\beta$ and $\theta$ bands exhibited lower $\hat{b}_{age}$ standard errors. In regions with approximately $n>200$, the $SE(\hat{b}_{age})$ values dropped below 0.0002. For $\alpha$, these same regions consistently showed a regression coefficient standard error of half the size, supporting the positive relationship between RBP($\alpha$) and age.

ROI-level confidence intervals also demonstrated that $\hat{b}_{age}$ standard errors varied with frequency band and sample size. Figure \ref{F:ROI_stats}B shows that in $\theta$ and $\gamma$ bands, regions are dominated by confidence intervals for $\hat{b}_{age}$ that contain 0, whereas the converse is true for $\delta$, $\alpha$ and $\beta$. 

With the exception of the $\theta$-band, these findings became robust in ROIs with $n \geq 291$. This suggests that equal and large sample sizes in all ROIs may lead to more uniform results and provides further confidence in the relationships we have identified between age and $\delta$, $\alpha$ and $\beta$ bands, and the lack of one between RBP($\gamma$) and age.

The age model returned some singular fits at the regional level (Figure \ref{F:ROI_stats}B). These regions typically had a low number of data points and a small number of hospitals.
We retained all models in all ROIs for completeness.

\section{Discussion} 

In this study, we considered how age, sex, and recording hospital, might impact RBP($\cdot$) in the setting of a normative icEEG map. To understand the effects of the variables, various LMMs were fitted, and it was concluded that sex and RBP($\cdot$) were not related in this cohort, whilst RBP($\cdot$) values present a moderate relationship with subject age and a notable relationship with the recording hospital. 

This work provides insight into what the trajectory of RBP($\cdot$) extracted from icEEG might look like in healthy lifespan. Additionally, it confirms that hospital site effects must be considered when modelling multi-centre data, whilst also demonstrating that sex perhaps need not be considered when analysing normative icEEG RBP($\cdot$).

\subsection{The relationship between hospital, sex, age and band power}

\subsubsection*{Hospital and band power}

In other neuroscience research concerned with normative mapping, such as neuroimaging, the multi-centre problem is widely recognised \citep{Hu2023, Keshavan2016, Jovicich2006}. It is common practice to model hospital site effects (or `batch' effects in neuroimaging literature) as random offsets in normative models \citep{Little2024, Hibar2018, Ge2024, Bethlehem2022, Kim2022}. Whilst it is impossible to capture all sources of hospital (or scanner) heterogeneity, there is some consideration of them in this field. Meanwhile, other neuroscience studies, such as those in EEG, do not typically consider hospital site effects.

For example, the RAM data set involves nine hospitals and is regularly used in academic work to produce results in icEEG \citep{Taylor22, Wang2023, Nozari2023, Kozma2024, Goldstein2019, Das2022}. Whilst the multi-centre challenge here has been recognised before by \cite{Miller2019-RAM}, to our knowledge, this and other studies have not statistically accounted for the multi-centre structure present. Our work reveals that, in the relative band power properties, the size of the hospital effect is frequency band specific, accounting for up to 30\% of the variation.

This demonstrates the necessity of hospital effect consideration, but also raises the question of what drives the differences. Supplementary \ref{S:HospEffects} considers the metadata we have available -- electrode type, age range and originating cohort (RAM/other) -- as possible contributors to recording hospital differences, but finds no systematic variation in these factors. Other technical aspects could be involved, such as recording settings or electrode manufacturer. Demographic possibilities include race, geography and the hospital's subject selection criteria.

Whilst both depth and grid/strip electrodes have differing advantages \citep{Taussig2015}, we found no evidence that electrode type was driving recording hospital variation (Supplementary \ref{S:HospEffects}). Past work found that both types were efficient for presurgical evaluation, and \cite{Frauscher2018} do not find any change in spectral density due to electrode type.

Our data did not contain information on subject ethnicity, however, this data should be collected in the future, to address the issue of past exclusionary practices in neuroscience \citep{Ricard2022, Li2022}. Previous research found race and marital status to be marginally associated with whether or not an individual proceeds to epilepsy surgery \citep{Berg2003}, so it is plausible such factors could also impact whether individuals undergo intracranial examination. 

Finally, we wish to highlight the dependency of the hospital site effect size on signal feature in our cohort (Table \ref{T:ICCR2}). For example, the hospital site effect was weakest in RBP($\alpha$). Investigating whether this is reproducible in other data sets is a further avenue of future research. There may also exist signal features which are more robust towards the effect of recording hospital.

In summary, although the multi-centre effect is under-explored in icEEG research, it is present in our work. Future work should improve on our own by collecting more metadata, to help identify what is driving the hospital site effect and to explain why different signal properties yield different results. 



\subsubsection*{Age and band power}

Moving to the fixed effects, our work produced convincing results in the $\delta$, $\alpha$ and $\beta$ bands when testing for a relationship between RBP($\cdot$) and age. Specifically, RBP($\delta$) decreased with subject age, while RBP($\alpha$) and RBP($\beta$) increased. More generally, we found an increase in RBP($\cdot$) with age in fast bands and a decrease in RBP($\cdot$) with age in slower bands.

These fixed effed results concur with previous literature on scalp EEG, although here the age range is reduced to children and adolescents, which found fast bands increased and the slow bands decreased in RBP($\cdot$) \citep{Gasser1988, Clarke2001}. Using the same age range, MEG power spectral density maps found an analogous relationship between power and age \citep{Ott2021}.

Looking at the full lifespan (7-84 years) using MEG, previous work found identical trends (a decrease in RBP($\delta$) and RBP($\theta$), and the converse for higher bands); however, the authors report slight changes around the sixth decade of life \citep{Gómez2013}. Since our maximum age is 18 years less at 66, this result still complements our findings. Further, a study using a small set of scalp electrodes across a large adult cohort found the same age trajectories as ours between 20-40 years, including the strongest decrease being in the $\delta$-band \citep{Hashemi2016}. However, they found that these effects diminished in older age ranges, which our choice of linear modelling does not facilitate.

Finally, there is evidence in the literature of a decrease in alpha band power with age -- \cite{Trondle2023} and \cite{Turner2023} report this across adulthood, and \cite{Whitford2007} in adolescence. However, these studies use absolute power rather than relative, in scalp rather than intracranial EEG, so this does not contradict our findings. \cite{Trondle2023} also argue that the decrease has been overestimated due to a lack of consideration of the aperiodic component. 

\subsubsection*{Sex and band power}

Our work did not find any evidence for a sex-RBP($\cdot$) relationship in this cohort. In the literature however, \cite{Ott2021} report sex differences in MEG power spectral density maps for children and adolescents in $\delta$, $\beta$ and $\alpha$. They do not report a main effect of sex in $\theta$ and $\gamma$, which are the same bands in which we found no notable relationship for either fixed effect. Further, \cite{Hashemi2016} report significant sex differences in total band power with age in an adult cohort using scalp EEG. 

Nevertheless, as with the evidence for the lifespan decrease in alpha power, both studies use a different, although related, feature and modality, which may facilitate the sex differences. Finding a difference in absolute power \citep{Hashemi2016}, but not relative power (our cohort) may suggest that the distribution of power is the same for men and women, but that the values differ. Alternatively, it is plausible that there is a genuine effect of sex on RBP($\cdot$) but that it is too small to detect in our cohort. 

In summary, the age trajectories we have found are generally in line with previous literature in similar areas, but a conclusion on the presence of sex differences remains unclear.

\subsection{Existing normative mapping work and how it relates to the ageing patterns}

Normative mapping is well-established in neuroscience research showing promising results in a range of modalities including MEG, scalp EEG and icEEG \citep{Niso2019, Owen2023, Janiukstyte23, Taylor22, Frauscher2018}. Evidence for the validity of normative maps has been studied, along with their temporal stability \citep{Rutherford2023, Wang2023, Janiukstyte23}. Generally, the patterns noted in neurophysiological normative maps complement each other, indicating their value. For example, many identify well-known trends such as $\alpha$ dominance in parietal and occipital regions \citep{Taylor22, Owen2023, Janiukstyte23, Frauscher2018, Groppe2013, Bosch-Bayard2020alpha}. 

The majority of normative mapping studies outlined consider their maps from a static viewpoint and did not incorporate potential age effects. In this study, we demonstrated that RBP($\cdot$) varies with age and that this variation is frequency band-specific. This is not surprising, as there is much previous work evidencing that healthy ageing \textit{does} impact brain activity in modalities such as MEG and scalp EEG \citep{Gomez2017, Gómez2013, Duffy1993}. Therefore we suggest, based on our findings, that future work no longer considers icEEG normative maps as a static snapshot and instead strives to incorporate dynamic features, such as age, into any study. 

\subsection{Clinical potential} 

Turning now to an application of normative maps, the comparison of individuals with epilepsy (or other disorders) to normalised healthy controls is a common one. It is gaining traction in the field of neurophysiology, using modalities which facilitate said controls such as scalp EEG and MEG \citep{Bosch-Bayard2020eegToolbox, Owen2023, Janiukstyte23, Owen2023clustering}. Similar work has been produced in the invasive icEEG setting; hence, it is plausible that the ageing trends demonstrated here could yield similar clinical potential \citep{Taylor22, Bernabei2022, Kozma2024}.

Whilst readily interpretable plots such as those in Figure \ref{F:brains}B summarise findings well, the core analysis is LMMs fitted at regional (Figures \ref{F:brains}A) or whole-brain levels (Table \ref{T:b_age_CIs}). Therefore, to localise pathological tissue in a clinical setting following an icEEG exam, we might compare an individual's age and RBP($\cdot$) in each frequency band to the corresponding regional regression line, quantifying differences, to determine whether their RBP($\cdot$) deviates from expected values at that age. Identifying regions with extreme deviations could guide the location of further examination. This particular pipeline would be similar to that suggested in \cite{Taylor22}. In the case of epilepsy, this could complement the standard seizure onset zone localisation techniques following an icEEG exam, without further work or procedures.


\subsection{Limitations and future directions}

The concept of `normative' data collected from individuals with epilepsy is perhaps contentious. These methods are defensible \citep{Rutherford2023} and have been applied in previous work \citep{Bernabei2022, Frauscher2018, Kalamangalam2020, Kozma2024, Taylor22, Wang2023}; however, epilepsy is increasingly defined as a network disorder \citep{Rayner2017, Kramer2012, Bernhardt2015}, which contrasts with the notion of delineating normal and abnormal tissue. In particular, previous work has shown that more complete resection of seizure onset regions is not associated with more favourable surgical outcomes \citep{Gascoigne2024}. Hence, it is arguable that the icEEG description of pathological tissue is complex and incomplete. 

Additionally, sample size impacted results despite our efforts in data collection (Section \ref{R:samplesize}), with high regional variation in number of subjects. In an ideal world, we would attain a similar number of subjects per region so they could be more accurately compared. In practice, however, this is not feasible due to some areas being more prone to pathology and therefore being more likely to have electrodes implanted -- for example, drug-resistant epilepsies in adults are common in the temporal lobe \citep{Bernhardt2019}. 

A further drawback is that our maximum age was only 66 years, so we did not have the full lifespan. This will always be difficult due to the risk of surgical operations on the elderly \citep{Grivas2006}, which limits the ability to directly compare our results to the many other modalities which extend to much older ages \citep{Gómez2013, CamCan2014,Hashemi2016, Duffy1993}. Arguably the minimum age of our cohort (4 years) also does not reflect the full lifespan, however, research has shown that children at younger ages have successfully undergone icEEG-guided epilepsy surgery and tolerated the invasive exam well \citep{Taussig2016}. Future data collection efforts could focus on expanding our age range at both ends. 

Similarly, the proportion of paediatric individuals in our cohort is low (Figure \ref{F:EDA}). In future work, we aim to collect more paediatric recordings and analyse this age range in isolation in an icEEG normative setting. This would provide insight into the effects of ageing during this key stage of life, as previously done for both scalp and MEG \citep{Ott2021, Gomez2017, Gasser1988, Clarke2001}.

Future work might also consider different frequency ranges. This study involves frequency content from 1-77.5 Hz, but higher frequency bands can contain pathological high-frequency oscillations (HFOs), which are thought to delineate epileptogenic tissue \citep{Zweiphenning2016, Zweiphenning2019}. Previous work has provided a normative map for HFOs obtained from icEEG \citep{Frauscher2018-HFO}. Hence, accounting for HFOs as in \cite{Kuroda2021} and extending the frequency range of our normative map could improve its clinical potential for identifying pathological areas of the brain. 

A further line of enquiry would be to increase the complexity of the model beyond \textit{linear} mixed models and consider the quadratic case or beyond. This has been done in other modalities, which found some frequency bands required non-linear models of band power over age, whilst other bands did not \citep{Gómez2013, Duffy1993}. Future work might investigate whether icEEG trajectories mirror those results, or if perhaps the optimal model complexity is not only frequency band specific, but regionally specific as well. 

Combining the discussions around frequency ranges and model complexity, there is also evidence that more granular frequency bands could affect the model selection process. For example, using healthy MEG data to determine how RBP($\cdot$) changes with age, \cite{Gómez2013} found that a linear regression model performed best for RBP(low-$\beta$), but a quadratic model is preferable for RBP(high-$\beta$). Analogous results were found in scalp EEG in the $\alpha$-band as well as the $\beta$-band (more complex models were required for the upper end of the band range) \citep{Gasser1988}.

Different features of EEG could also be considered, in particular, absolute band power as an alternative to the relative band power used here. This would allow for direct comparison to the aforementioned work which finds sex differences in absolute power \citep{Hashemi2016}. However, a previous EEG study investigating age, sex and band power found that absolute power failed to find significant lifespan effects in all bands except $\delta$, whilst relative power succeeded, suggesting further investigation using absolute power is unnecessary \citep{Clarke2001}.

\subsection{Conclusions} 

In conclusion, our results suggest that whilst recording hospital and age have some impact on normative icEEG RBP($\cdot$) in this cohort, sex does not. Our work might be considered the first attempt to study the relationship between these variables and icEEG in a normative setting. Our results highlight the importance of accounting for the heterogeneity in icEEG data by including covariates such as age and recording hospital (where applicable) in future normative mapping work. 

Sample size is a key discussion point of this study, being both a strength and limitation of our work. Whilst we have collected one of the largest icEEG datasets to date, some of our results are limited by number of subjects per region. Future work could strive to address this, whilst also considering the impact of non-linear models, or the effects of using a different EEG feature. 

We propose that the dynamic nature of normative mapping should be acknowledged in future icEEG research, rather than only considering a static viewpoint that does not account for variables such as age. Further, multi-centre work needs to investigate and model the impact of using data from several recording hospitals to ensure the accurate interpretation of any results. 

\section{Data and code availability}

The preprocessed, normative data table including RBP($\cdot$) values, along with code used for modelling and to produce figures, we be made available at
https://github.com/H-Woodhouse/norm-ieeg-age-sex-hospital upon acceptance.

The data from RAM hospitals is publicly available at https://memory.psych.upenn.edu/RAM. Due to data sharing agreements, the raw icEEG for the remaining hospitals is not available. However, the CNNP lab will be publishing a data release paper containing this in the future. 

\section{Author Contributions}

\begin{itemize}
    \item Conceptualisation: YW HW
    \item Methodology: YW HW PNT GMS
    \item Analysis: HW
    \item Data Acquisition: HW BD JSD JM KZ AC MT FM CP MAH GMI ED NMW RA PNT YW
    \item Data processing: HW GH CS CK FT GMS PNT YW
    \item Writing - Original Draft: HW
    \item Writing - Review \& Editing: HW GMS JSD AC MT FM CP MAH GMI NMW PNT YW
    \item Supervision: YW
\end{itemize}

\section{Funding}

P.N.T. and Y.W. are both supported by UKRI Future Leaders Fellowships (MR/T04294X/1, MR/V026569/1). H.W. and C.S. are supported by the Engineering and Physical Sciences Research Council (EP/L015358/1); J.S.D. is supported by the NIHR UCLH/UCL Biomedical Research Centre. C.K. is supported by Epilepsy Research UK. R.A. is supported by the Clinical and Translational Science Award, Grant UL1TR002373, NIH/NCATS. G.M.I. was funded by the Abe Bresver Chair in Functional Neurosurgery at the Hospital for Sick Children.

\section{Declaration of Competing Interests}

None of the authors have any conflict of interest to disclose. 

\section{Acknowledgements}

We thank members of the Computational Neurology, Neuroscience \& Psychiatry Lab (www.cnnp-lab.com) for discussions on the analysis and manuscript. We thank our coauthors for their data contribution and feedback on the manuscript. 




\newpage

\renewcommand{\thesubsection}{S\arabic{subsection}}
\makeatletter
\def\@seccntformat#1{\@ifundefined{#1@cntformat}%
   {\csname the#1\endcsname\quad}
   {\csname #1@cntformat\endcsname}}
\newcommand\section@cntformat{}     
\makeatother
\setcounter{section}{0}

\renewcommand{\thefigure}{S\arabic{figure}}
\setcounter{figure}{0}

\renewcommand{\thetable}{S\arabic{table}}
\setcounter{table}{0}

\section{Supplementary Material}

\subsection{Algorithmic detection of noisy or faulty channels}
\label{S:BadChannels}

Channels were removed based on the five criteria given below. American hospitals were run separately from the European hospitals and Beijing, due to the line noise difference.

\begin{enumerate}
    \item Exclude channels based on channel details data. If channels are known to be resected, within structurally abnormal tissue or within the seizure onset zone, they are removed. This is determined by resection masks or clinical reports.
    \item Exclude algorithmically detected noisy channels by finding channels with outlier signal range and/or variance relative to the other channels.
    \begin{itemize}
        \item Two rounds of detection are performed; the first round is before preprocessing with (by default) less stringent detection thresholds, and the second is after basic preprocessing.
        \item The first round uses a threshold for outlier detection of 16, for both signal variance and range.
        \item Before the second round, common average is applied to the icEEG data.
        \item A 4$^{th}$ order band pass filter is applied between 0 and 100~Hz. 
        \item A notch filter is applied to eliminate the location-specific line noise (50~Hz for European hospitals and Beijing Tiatian Hospital and 60~Hz for American hospitals).
        \item The threshold for outlier detection based on signal range and variance in the second round following preprocessing is 12. 
    \end{itemize}
    \item Exclude channels previously marked as `bad'. These are the channels noted as noisy/faulty by visual inspection.
    \item Remove channels which are missing channel details.
    \item Remove channels without mapping/localisation to an ROI.
\end{enumerate}

\subsection{Available channel details by hospital site}
\label{S:ChannelDetails}

Three additional channel features were recorded if available through clinical reports or resection masks, namely, whether the channel was 1) within the seizure onset zone, if determined 2) located in structurally abnormal tissues such as lesions or, 3) resected during surgery. If a channel fell into any of these categories, it was removed from analysis. Table \ref{S-tab:contacts} demonstrates how many subjects (per hospital site) had these variables in their reports. 

\renewcommand\theadalign{tc}
\renewcommand\theadfont{\bfseries}
\begin{table}[h]
    \centering
    \begin{tabular}{|l|c|ccc|}
       \cline{2-5}
        \multicolumn{1}{c|}{} & \multirow{2}{*}{\thead{Total\\subjects}} & \multicolumn{3}{c|}{\thead{Number of subjects with\\information on:}}\\
        \multicolumn{1}{c|}{}  & & SOZ &  \makecell{Structural\\abnormality} & Resection \\
       \hline
       Beijing Tiantan Hosp.          & 32  & 32 &  0  & 28\\
       Columbia University*           & 1   & 1  &  1  & 0 \\
       Dartmouth University*          & 26  & 26 &  26 & 0 \\
       Emory University*              & 34  & 34 &  34 & 0 \\
       Great Ormond St. Hosp.         & 15  & 15 &  0  & 13\\
       Jefferson Hosp.*               & 41  & 41 &  41 & 0 \\
       Mayo Clinic*                   & 35  & 35 &  35 & 0 \\
       NINDS*                         & 16  & 16 &  16 & 0 \\
       SickKids                       & 28  & 0  &  0  & 0 \\
       University of Pennsylvania*    & 29  & 29 &  29 & 0 \\
       University College London Hosp.& 107 & 102&  0  & 106\\
       University of Iowa Hosp.       & 84  & 22 &  0  & 51\\
       University of Washington*      & 2   & 2  &  2  & 0 \\
       University of Wisconsin-Madison& 21  & 21 &  0  & 0 \\
       UT Southwestern*               & 31  & 31 &  31 & 0 \\
       \hline
    \end{tabular}
    \caption{For each hospital site, the number of subjects with information on which electrode contacts are: within SOZs, within structural abnormalities, resected.}
    \label{S-tab:contacts}
\end{table}

\subsection{Robustness to parcellation choice}
\label{S:ParcellationValid}

Our main results use the coarsest scale parcellation, scale 36, with data mirrored across hemispheres for sample size reasons. Here, we repeat Figure \ref{F:brains}B again in a finer-grained parcellation, the scale 60 atlas. Here, the data has \textit{not} been mirrored.

Due to a decrease in sample size per region, more regions produce a singular fit. Using the scale 36 atlas, as shown in Figure \ref{F:ROI_stats}B, there are a total of 9 singular fits out of 190 total fits (38 ROIs $\times$ 5 bands). When using the scale 60 atlas, without mirroring, there are 61 singular fits out of 610 total fits (122 ROIs $\times$ 5 bands), which is a significantly larger portion -- approximately double. This is due to the LMM fitted in each ROI being too complex for the size of the data at this finer-grained parcellation. For this reason, we did not consider any finer-grained atlases. 

\begin{figure}[h]
    \centering
    \includegraphics[width=\textwidth]{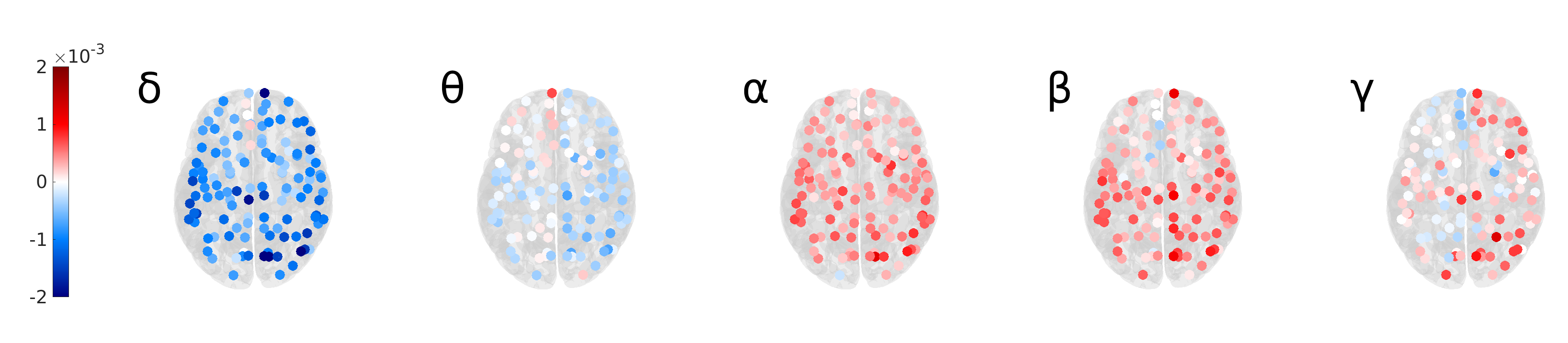}
    \caption{Values of $\hat{b}_{age}$ from the age model implemented at the region-level for the scale 60 atlas. Values are shown for each ROI and each frequency band of interest. The colour scale is symmetric and fixed across frequency bands with blue representing negative regression coefficients and red representing positive ones. The original data has been used so results are not reflected across the midline.}
    \label{S-fig:brainsROI2}
\end{figure}

Nevertheless, as shown in Figure \ref{S-fig:brainsROI2}, using the scale 60 atlas, results are reasonably symmetric in $\delta$, $\alpha$ and $\beta$. Any deviations from symmetry/expected results, tend to lie near the midline, whose regions have lower sample sizes than lateral ones. The $\theta$ and $\gamma$ bands do not exhibit symmetry, but as outlined in Section \ref{R:AgeModelBest}, their relationship with age was weak.

We conclude that while (for this cohort) the model under consideration is only suitable for the parcellation applied in the main paper, symmetry holds to a satisfactory level when considering one, finer-grained parcellation. 

\subsection{Validity of mirroring data in our parcellation}
\label{S:PoolingValid}

In Section \ref{M:pooling} we described the mirroring of symmetric regions in order to increase the sample size in each region. The only instance where we mirror across the midline for visualisation is found in Figure \ref{F:brains}B. Here we repeat that figure, without mirroring symmetric regions, to demonstrate that the main results are unchanged. Note that, due to the reduction in sample size per ROI using the full data, there are 28 singular fits, treble the number attained when using mirrored data. 

\begin{figure}[h]
    \centering
    \includegraphics[width=\textwidth]{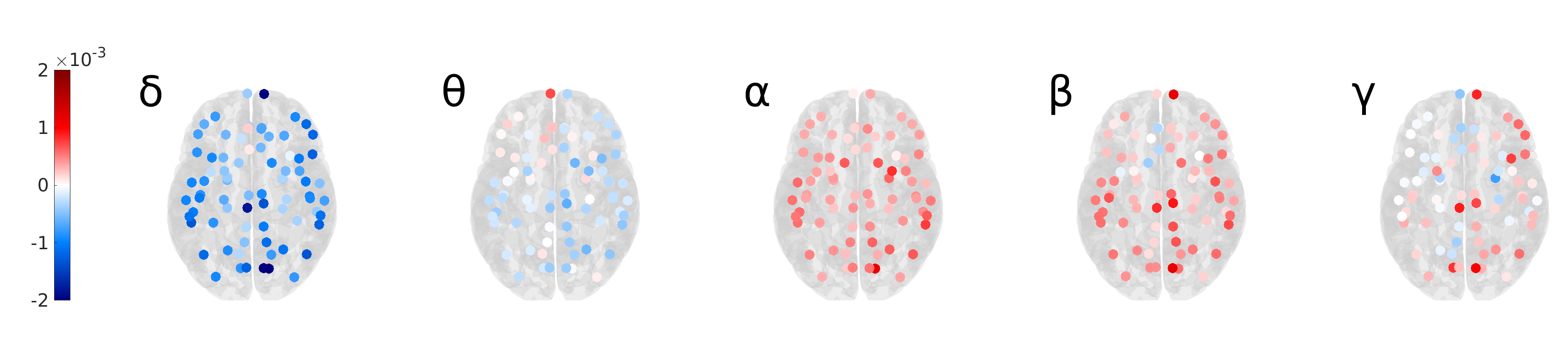}
    \caption{Values of $\hat{b}_{age}$ from the age model implemented at the region-level for the scale 36 atlas. Values are shown for each ROI and each frequency band of interest. The colour scale is symmetric and fixed across frequency bands with blue representing negative regression coefficients and red representing positive ones. The original data has been used so results are not reflected across the midline.}
    \label{S-fig:brainsUnpooled}
\end{figure}

Figure \ref{S-fig:brainsUnpooled} demonstrates that results are symmetric when using the original data in bands where we have identified a relationship between RBP($\cdot$) and age, and the overarching trends are unchanged. We see a decrease in RBP($\delta$) and RBP($\theta$) with age, an increase in RBP($\alpha$) and RBP($\beta$), and weak results for RBP($\gamma$). There is no spatial specificity in the results, and trends are weaker in the $\theta$ and $\gamma$ bands. This is all in line with results presented in Result \ref{R:age&space}.

\subsection{Normative values are not correlated with age of epilepsy onset}
\label{S:AgeOnset}

After implementing the age model in all frequency bands at the ROI level using the mirrored data, we tested for correlation between model residuals and age of epilepsy onset. 

\begin{figure}[h!]
    \centering
    \includegraphics[width=0.7\textwidth]{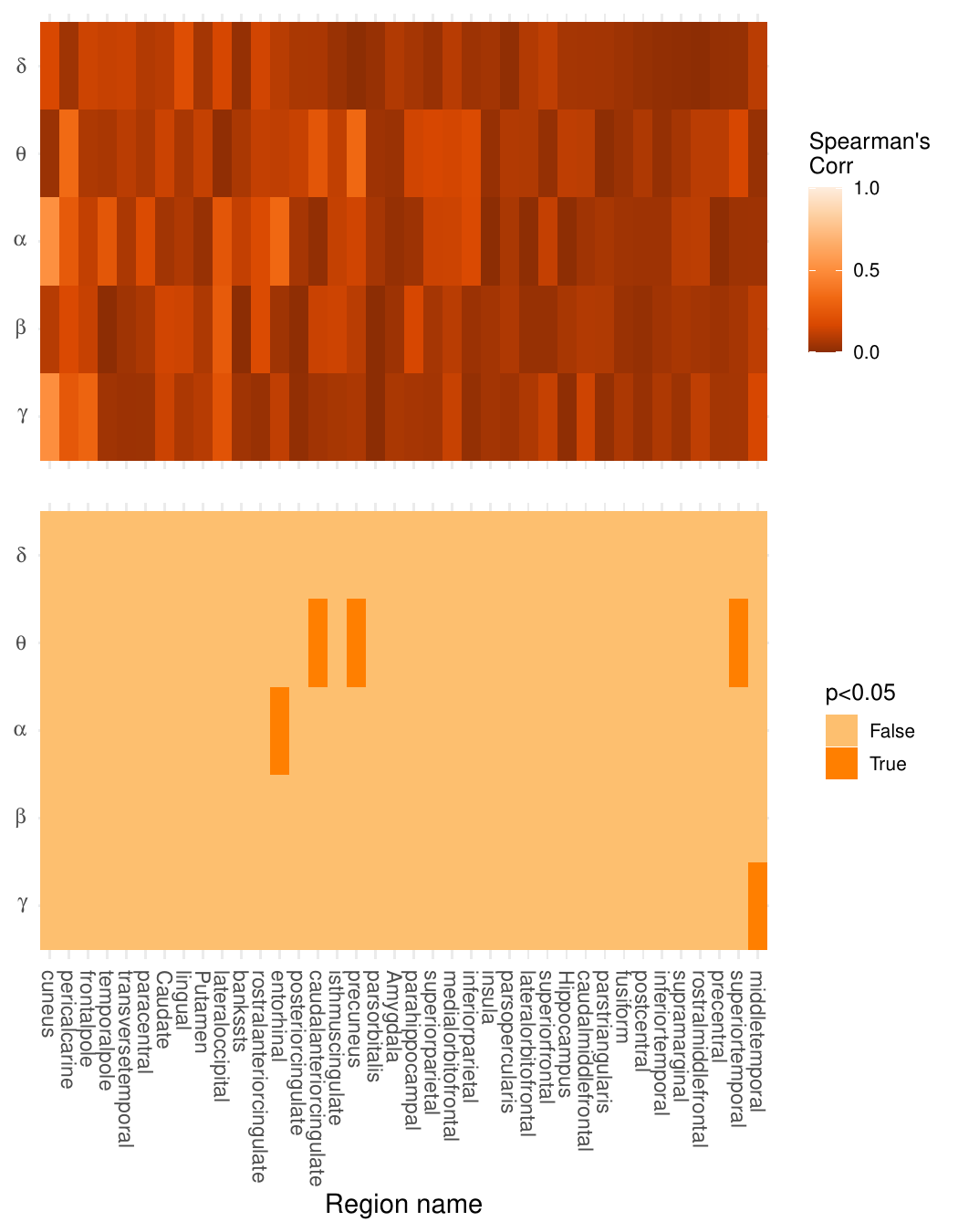}
    \caption{Heatmaps representing Spearman's correlation test applied to the residuals of the age model and age of epilepsy onset, in every frequency band and ROI. Absolute correlation coefficients (top) are on a gradient colour scale with near-white representing high correlation and dark red representing low values. Corresponding $p$-values (bottom) have a binary indicator of significance at the 5\% level. We have not applied a multiple-comparison correction, but the rate of detection is in agreement with and below what would be expected at 5\% level (5 detections out of 190 tests). The regions are ordered by number of subjects, from low to high.}
    \label{S-fig:age_of_onset}
\end{figure}

Not every subject has age of onset data, so we fit the age model to all data, then applied Spearman's test using only complete pairs. Figure \ref{S-fig:age_of_onset} shows the absolute value of the correlation estimate in each ROI and frequency band (top). Near-white values demonstrate strong correlation (in either direction) whilst dark red values indicate no, or weak, correlation. The $p$-value plot (bottom) on the same axes shows a dark orange when $p<0.05$. It shows a light orange if the converse is true. The $x$-axis is ordered from lowest to highest number of subjects per region.

Figure \ref{S-fig:age_of_onset} is dominated by low correlations and non-significant $p$-values, indicating no relationship between age of onset and normative band power values, and further supporting that our data are representative of normative, rather than pathological activity. 

\subsection{No evidence of a non-linear relationship between RBP(\texorpdfstring{$\cdot$}{}) and age}
\label{S:LinearValid}

\begin{figure}[h!]
    \centering
    \includegraphics[width=0.85\textwidth]{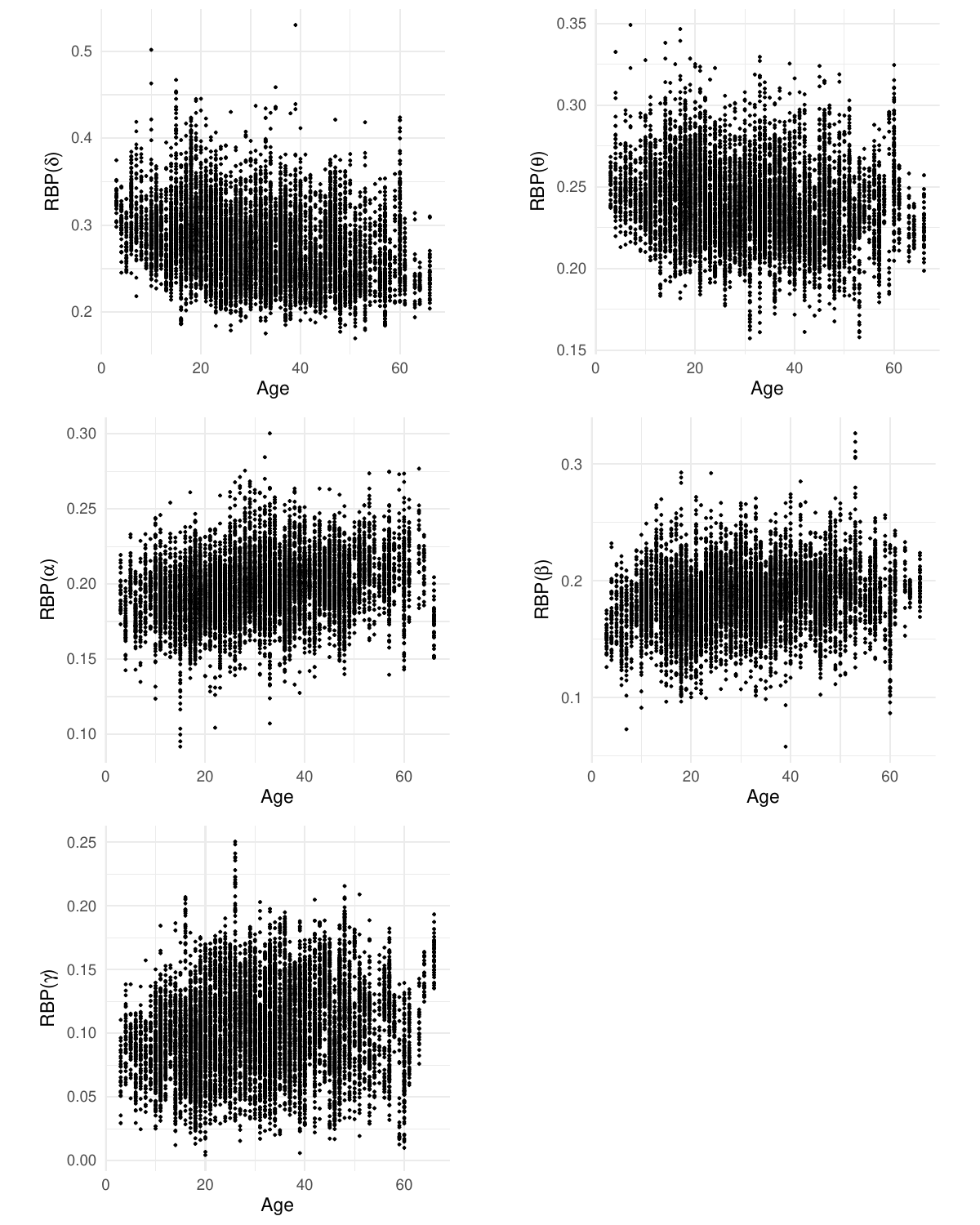}
    \caption{Scatter plots of RBP($\cdot$) against age in each frequency band, using the original data at the whole-brain level}
    \label{S-fig:scatterplot}
\end{figure}

We decided a linear mixed model would be suitable for this work (Section \ref{M:optLMM}) as we did not see strong evidence that the relationship between RBP($\cdot$) and age was non-linear. Figure \ref{S-fig:scatterplot} shows all data at the whole-brain level, and presents a generally linear trend in each frequency band.

\subsection{Consideration of an interaction term between age and sex}
\label{S:interaction}

During the model selection process in Section \ref{M:optLMM}, we additionally considered a fifth covariate structure involving an interaction term, giving a final model for each frequency band,
$RBP(\cdot) \sim Age + Sex + Age*Sex + (1|Hospital)$.

During the model selection process, $\delta$, $\beta$ and $\gamma$ returned non-zero confidence intervals on the interaction term, however, model evaluation statistics did not unanimously select the interaction model, and it performed similarly to the simpler, optimal choice in each case. In the $\alpha$-band the interaction was not significant. The $\theta$-band selected the interaction model, however as determined throughout our results, RBP($\theta$) has a weak relationship with our covariates. For these reasons, we didn't consider the interaction term in the main text. 

To demonstrate the lack of value of an interaction term further down the pipeline, we repeat Table \ref{T:ICCR2} here with the interaction model included. Looking at the $R^2_m$ columns, the inclusion of an interaction term adds a maximum of 0.18\% to the variation in RBP($\cdot$) explained by our fixed effects structure. In the $\theta$-band where the interaction model is preferred, the response variance explained by fixed effects is $<1\%$. Hence an interaction term was not considered within our main results. 

\begin{table}[h]
    \setlength{\tabcolsep}{14pt}
    \centering
    \begin{tabular}{|c|cc|cc|cc|cc|}
    \cline{2-9}
    \multicolumn{1}{c|}{} & \multicolumn{2}{c|}{\makebox[0pt]{Interaction model}} &\multicolumn{2}{c|}{Full model} & \multicolumn{2}{c|}{Age model} & \multicolumn{2}{c|}{Sex model}\\
    \multicolumn{1}{c|}{} & $R^2_m$  & $ICC$ & $R^2_m$  & $ICC$ & $R^2_m$ & $ICC$ & $R^2_m$ & $ICC$ \\
    \hline
       $\delta$  & 4.46 & 15.26 & 4.28  & 15.23  & \cellcolor{delta}4.27 & \cellcolor{delta}15.27  & 0.00 & 19.42 \\
       $\theta$  & \cellcolor{theta}0.70 & \cellcolor{theta}15.10 & 0.60  & 15.17 & 0.23 & 14.88 & 0.41 & 16.25 \\
       $\alpha$  & 8.07 & 5.16 & 8.04  & 5.09  & \cellcolor{alpha}7.97 & \cellcolor{alpha}5.13  & 0.00 & 5.88  \\
       $\beta$   & 2.71 & 6.59 & 2.59  & 6.58  & \cellcolor{beta}2.58 & \cellcolor{beta}6.58  & 0.00 & 7.93  \\
       $\gamma$  &0.37 & 32.55 &  0.25  & 32.59 & 0.01 & 32.16 &  \cellcolor{gamma}0.24 & \cellcolor{gamma}32.25 \\
    \hline
    \end{tabular}
    \caption{$R^2_m$ and $ICC$ values (measured in \%) for the interaction model, the full model, the age model and the sex model. $R^2_m$ represents the proportion of variation in RBP($\cdot$) explained by the fixed effect(s) of that model. The $ICC$ represents the proportion of the variance explained by the grouping structure, namely recording hospitals. The optimal covariate subset, as determined by a standard model selection process, is highlighted for each frequency band.}
    \label{S-tab:ICCR2}
\end{table}

\subsection{Hospital site effects across all frequency bands and hospitals}
\label{S:HospEffects}

In Figure \ref{F:hosp_effects}, we used one frequency band and three hospitals to demonstrate the hospital effect visually, overlaid on the normative data points. Here, we fit the age model on the whole brain level to demonstrate the varying hospital intercepts across all hospital sites and frequency bands, to highlight the importance of the consideration of such effects.

Figure \ref{S-fig:intercepts_offsets} shows (for each frequency band) the deviation of each hospital's intercept from the population intercept, along with a standard 95\% confidence interval. In other words, each hospital's deviation from the mean population RBP($\cdot$). Additionally, metadata has been included on the left of the plot, indicating which hospitals originated from the RAM database, which hospitals used only depth electrodes, and which hospitals had only paediatric, or only adult subjects. On the $y$-axis, hospitals are ordered by deviation from the overarching intercept in that band. The $x$-axis range is fixed across figures, demonstrating how some frequency bands show greater variation across recording hospitals. 

Columbia University and the University of Washington consistently display notably large confidence intervals as a result of their low sample size. Whilst some hospitals fall above or below population means in each band, there are no hospitals which are systematically different across all five subplots. Further, there are no groupings, such as only paediatric hospitals, which show a consistent pattern across signal properties. 

Hence, we do not have sufficient metadata to determine what is driving the large, varied hospital site effects, it does not appear to be any of the variables present in our cohort. Further investigation is required to determine the source of the recording hospital differences in RBP($\cdot$).

\begin{figure}[h!]
    \centering
    \begin{subfigure}{\textwidth}
        \includegraphics[width=\linewidth]{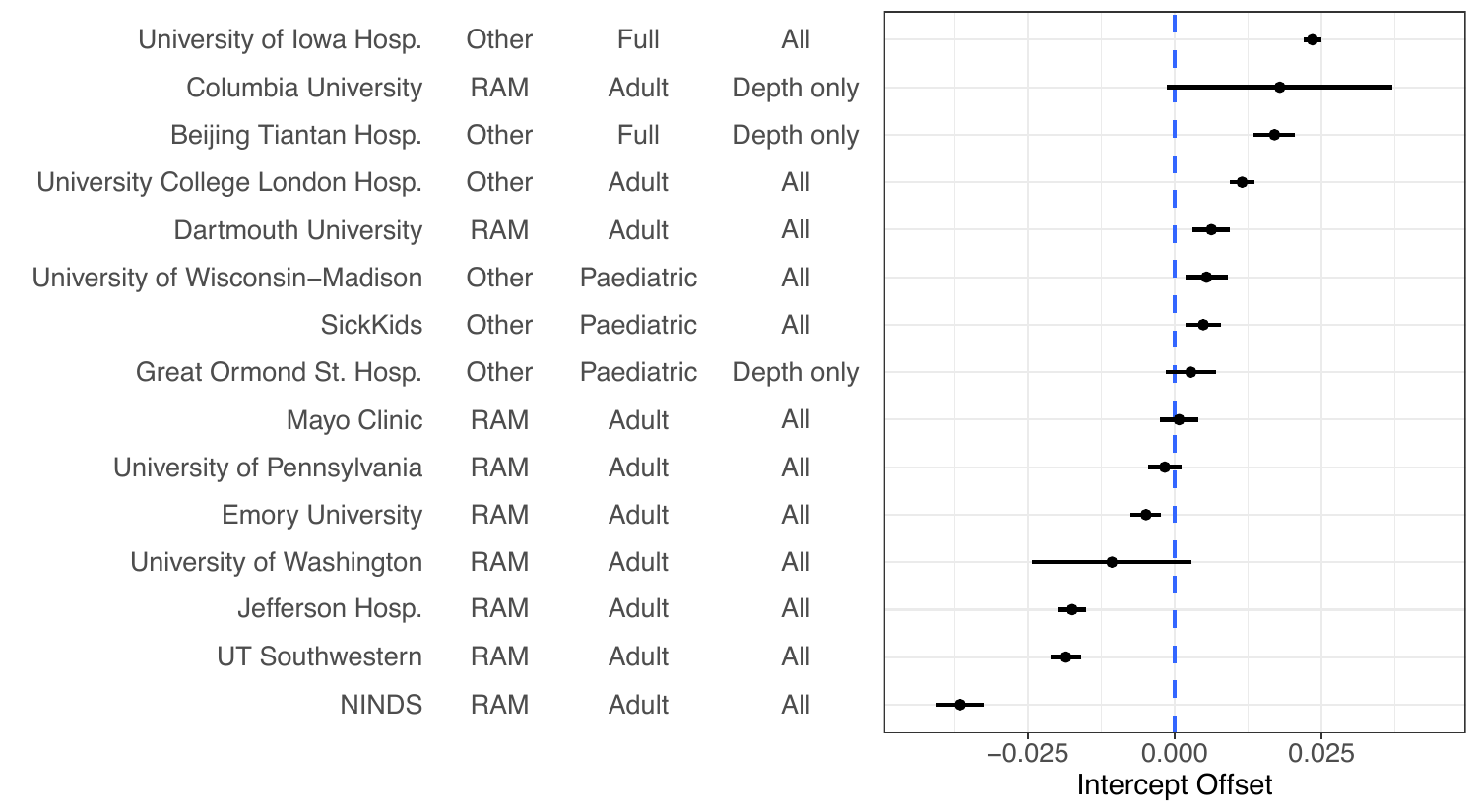}
        \subcaption{$\delta$-band}
        \vspace{1.5cm}
    \end{subfigure}
    \begin{subfigure}{\textwidth}   \includegraphics[width=\linewidth]{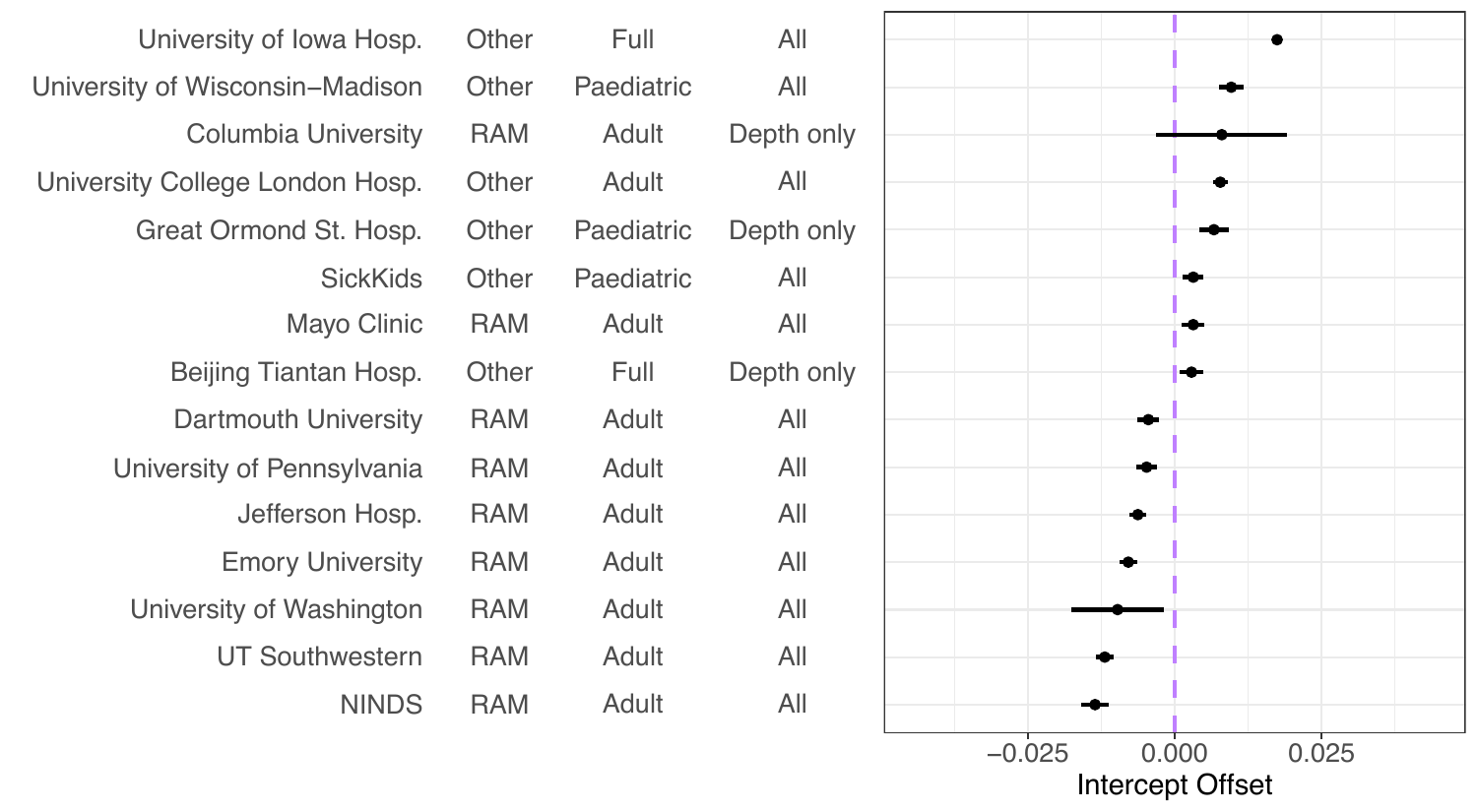}
        \subcaption{$\theta$-band}
    \end{subfigure}
\end{figure}

\begin{figure}[H]\ContinuedFloat
    \centering
    \begin{subfigure}{\textwidth}
    \includegraphics[width=\linewidth]{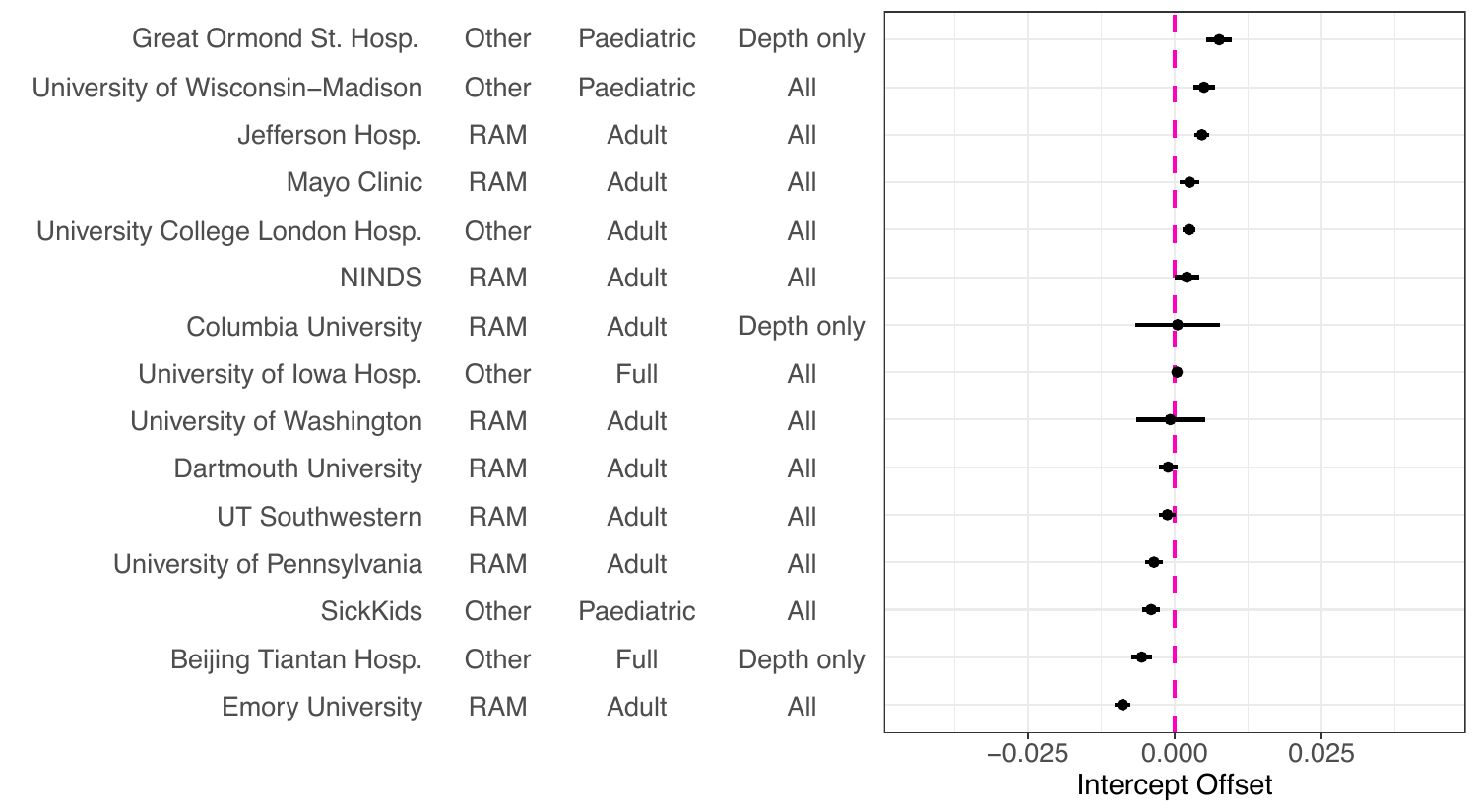}
    \subcaption{$\alpha$-band}
    \vspace{1.5cm}
    \end{subfigure}
    \begin{subfigure}{\textwidth}
    \includegraphics[width=\linewidth]{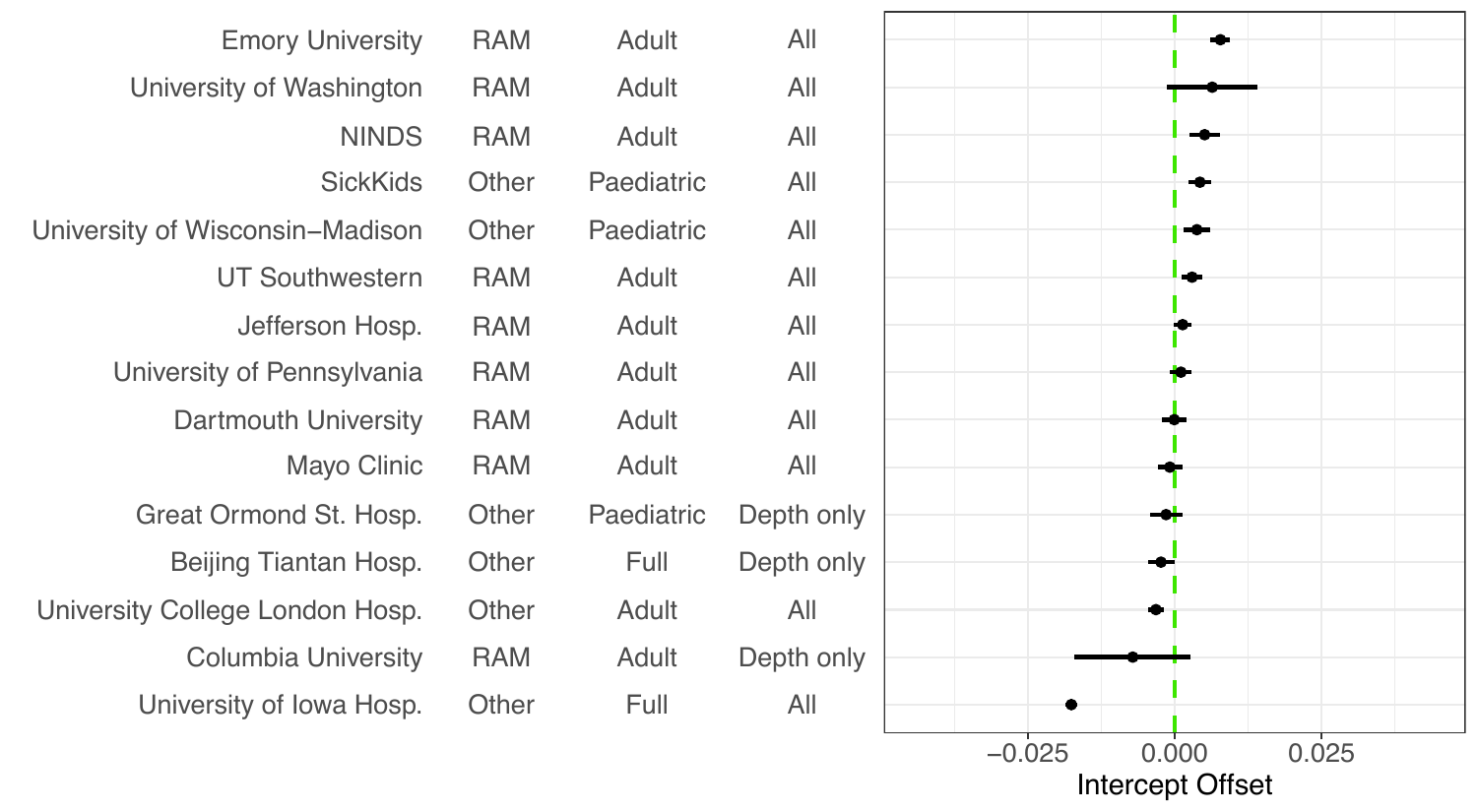}
    \subcaption{$\beta$-band}
    \vspace{1.5cm}
    \end{subfigure}
\end{figure}

\begin{figure}[h!]\ContinuedFloat
    \centering
    \begin{subfigure}{\textwidth}
    \includegraphics[width=\linewidth]{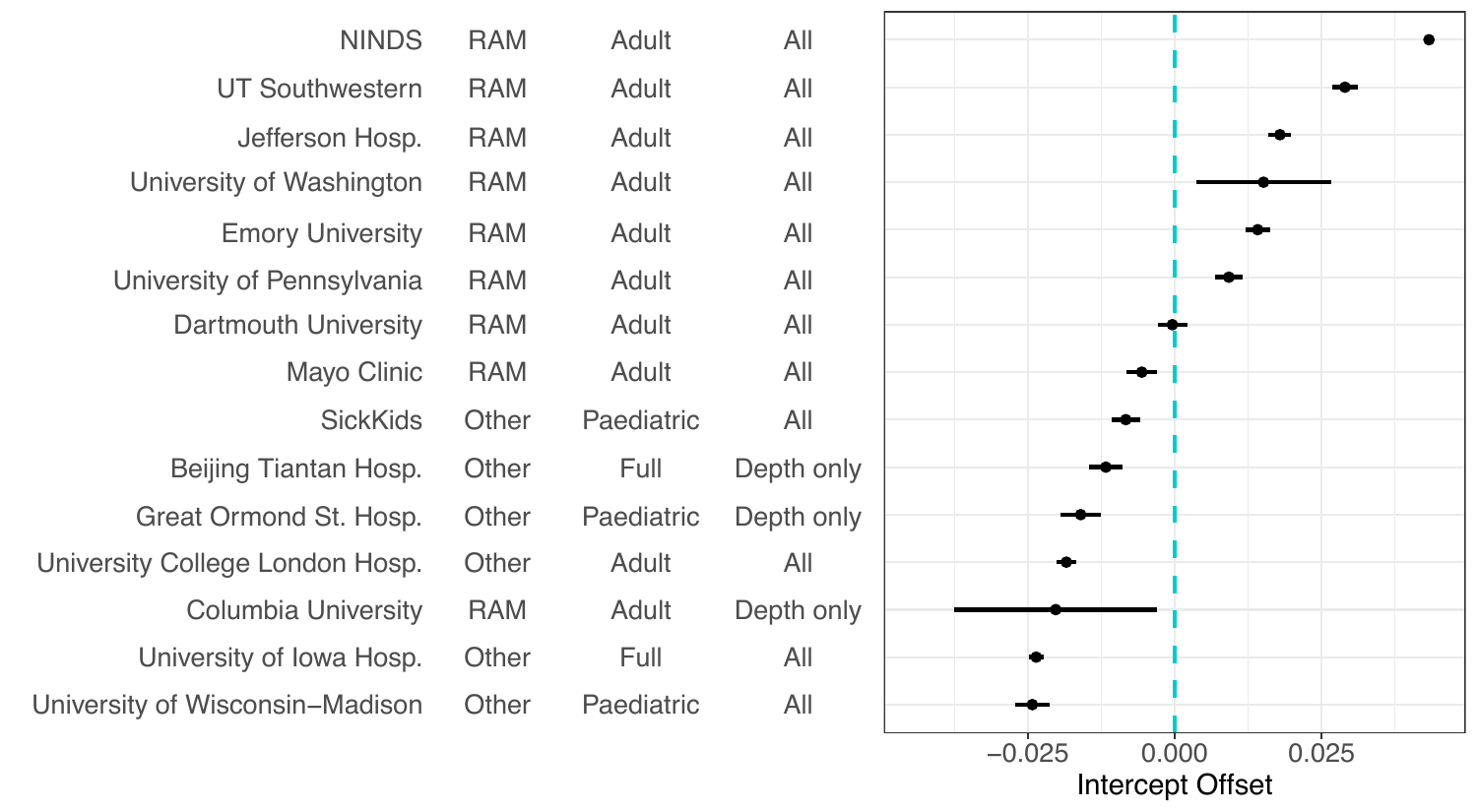}
    \subcaption{$\gamma$-band}
    \end{subfigure}
   \caption{For each frequency band \textbf{a)-e)}: Plots showing each recording hospital's deviation from the population intercept when fitting the age model at the whole brain level. The $x$-axis is fixed at a range of $(-0.05,0.05)$ for all bands. The hospitals are ordered on the $y$-axis by largest positive offset from the RBP($\cdot$) population mean, to largest negative offset. Three metadata columns on the left of the plot show three grouping variables. Namely 1) if the hospital originated from the RAM dataset, 2) if the hospital comprises only paediatric subjects, only adults, or the full age range, 3) if the retained subjects from the hospital had only depth electrode implantation, or whether there were some individuals with grid/strip electrodes as well.}
    \label{S-fig:intercepts_offsets}
\end{figure}

\subsection{Age distributions at ROI-level}
\label{S:ROIages}

Results \ref{R:age&space} and \ref{R:samplesize}, employ the age model at the regional level. Whilst we discuss the number of subjects per model, this could be misleading in this context if, for example, a highly populated ROI only consisted of a very narrow age range. Such a scenario would undoubtedly impact any results surrounding $\hat{b}_{age}$. 

\begin{figure}[h!]
    \centering
    \includegraphics[width=0.85\textwidth]{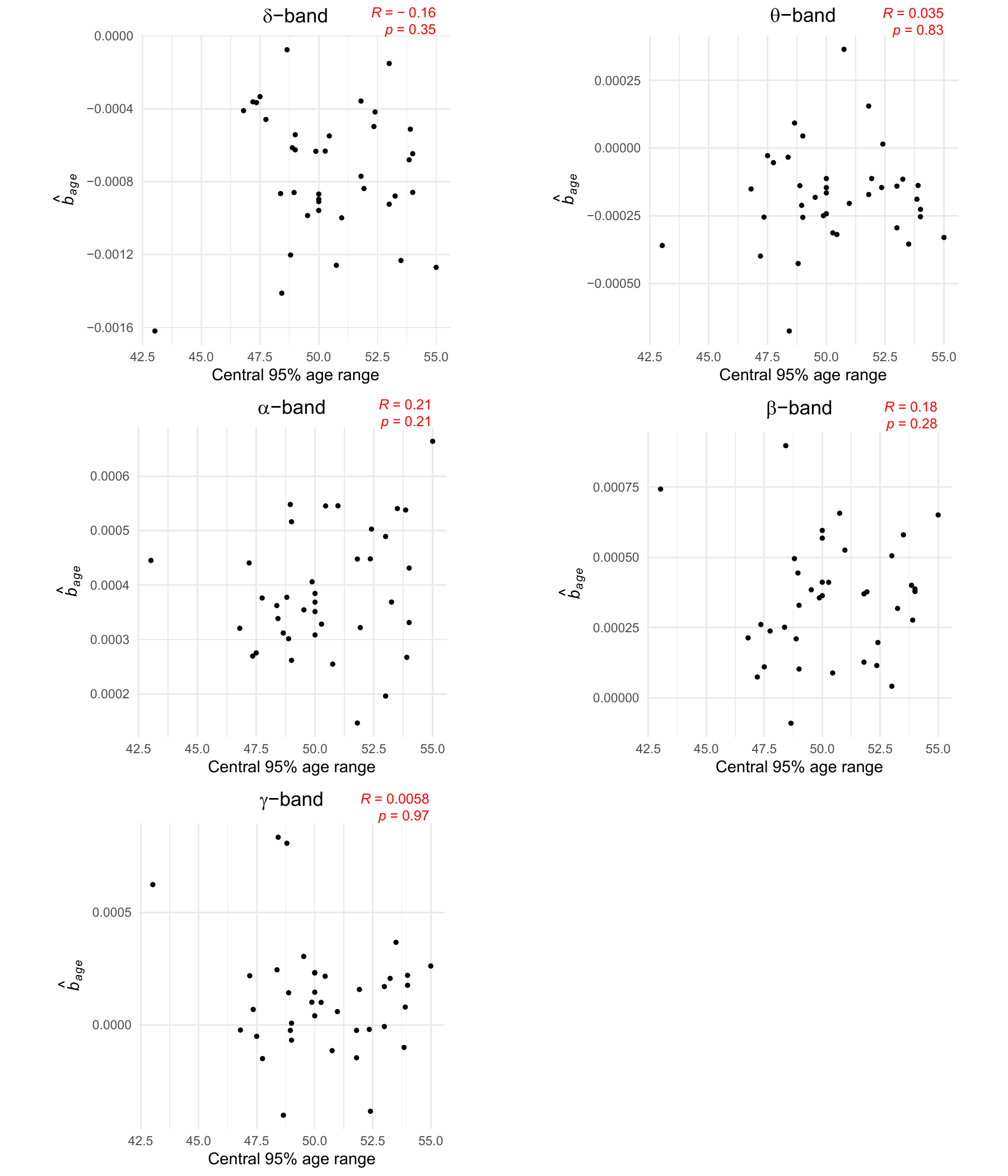}
    \caption{Values of $\hat{b}_{age}$ from the age model implemented at the region-level for the scale 36 atlas plotted against the central 95\% percentile range of subject ages in that region. Results are given for all frequency bands with Spearman's correlation coefficient and resulting $p$-value in the top-right corner.}
    \label{S-fig:age_b_corr}
\end{figure}

Hence, we have calculated the 2.5th and 97.5th percentiles to give the 95\% central age range in each region, and plotted this against the regions $\hat{b}_{age}$ value. This has been repeated for all five frequency bands and the result can be seen in Figure \ref{S-fig:age_b_corr}.

All ROIs have a minimum 2.5-97.5th percentile age range of 43 years, with about half the regions reaching 50 years or more. Additionally, in all five frequency bands, there is no correlation between $\hat{b}_{age}$ and the distribution of ages, as confirmed by Spearman's method. Therefore, there is a reasonable distribution of subject ages in each ROI, and there is no evidence that this distribution is influencing regional results. It is acceptable to discuss sample size of regions as the influencing factor in the relevant results. 

\newpage
\bibliography{ref}

\end{document}